\documentclass{jkas}

\def\year{2015} 
\def\month{December} 
\def\volume{48} 
\def\issue{6} 
\def\beginpage{1} 
\def\endpage{99} 
\def\received{November 30, 2015} 
\def\accepted{December 16, 2015} 
\date{Received \received; accepted \accepted}

\usepackage{flushend} 
\def\asec{$^{\prime\prime}$}
\def\lum{erg s$^{-1}$}
\def\perhz{Hz$^{-1}$}
\def\solmass{$M_\odot$}
\def\peryr{yr$^{-1}$}

\title{
Study of Milli-Jansky Seyfert Galaxies with Strong Forbidden
High-Ionization Lines Using the Very Large Array Survey Images
}

\author{Dharam~V.~Lal}

\affil{National Centre for Radio Astrophysics (NCRA--TIFR), Pune University Campus,
Post Box 3, Ganeshkhind P.O., Pune 411007, India; \email{dharam@ncra.tifr.res.in}}

\def\corrauthor{
D. V. Lal
}

\def\runningauthor{
Lal
}

\def\runningtitle{
Radio Properties of Forbidden High-Ionization Line Emitting Seyfert Galaxies
}

\def\keywords{
galaxies: active --- galaxies: jets --- galaxies: nuclei ---
galaxies: Seyfert --- galaxies: structure --- radio continuum: galaxies
}

\def\abstracttext{
We study the radio properties at 1.4 GHz of Seyfert galaxies with
strong forbidden high-ionization lines (FHILs),
selected from the Sloan Digital Sky Survey -- a large-sized sample
containing nearly equal proportion of diverse range of
Seyfert galaxies showing similar redshift distributions
compiled by \citet{GMW2009} using the Very Large Array survey images.
The radio detection rate is low, 49\%, which is lower than
the detection rate of several other known Seyfert galaxy samples.
These galaxies show low star formation rates and the radio emission is dominated
by the active nucleus with $\le$10\% contribution from thermal emission,
and possibly, none show evidence for relativistic beaming.
The radio detection rate, distributions of radio power, and
correlations between radio power and line luminosities or X-ray luminosity for
narrow-line Seyfert~1 (NLS1), Seyfert~1 and Seyfert~2 galaxies are
consistent with the predictions of the unified scheme hypothesis.
Using correlation between radio and [O\,III]\,$\lambda$\,5007\,\AA\
luminosities, we show that $\sim$8\% sample sources are radio-intermediate and
the remaining are radio-quiet.
There is possibly an ionization stratification associated with clouds
on scales of 0.1--1.0~kpc, which have large optical depths at 1.4\,GHz,
and it seems these clouds are responsible for free--free absorption of
radio emission from the core; hence, leading to low radio detection rate
for these FHIL-emitting Seyfert galaxies.
}

\begin{document}
\jkashead 


\section{Introduction}
\label{sec:intro}

Spiral galaxies, having bright star-like nuclei covering a wide
range of ionization, are known as Seyfert galaxies.
They are generally classified into type~1 and type~2,
which depends only on relative nuclear emission-line widths
\citep{KW74}.
The Seyfert~2 galaxies have relatively narrow permitted hydrogen lines and
narrow forbidden lines,
whereas Seyfert~1 galaxies have broad permitted hydrogen lines
and narrow forbidden lines.
The widths of narrow lines and of broad lines in terms of full width
at half maximum (FWHM) are ~$\approx$~300--1,000~km~s$^{-1}$ and
$\ge$~1,000~km~s$^{-1}$ for type~2 and type~1, respectively.
An orientation based unified scheme
is often used in explaining the
classification of a Seyfert galaxy, where
the Seyfert~2 represent edge-on source and
the Seyfert~1 galaxy is its pole-on equivalent
\citep[e.g.,][]{antonucci,Urry,Deoetal2007,Netzer2015}.
Observational evidence for and against this unified scheme
hypothesis exist on various scales and in all wavebands
\citep[see][for a brief summary]{Singhetal2013}.
In all these studies, sample selection is the key; for example,
biases against obscured sources, and biases towards dusty sources
are suggested for the Optical-/UV-selected samples; whereas
biases towards sources with higher level of nuclear star formation are suggested
for the infrared-selected sources \citep{Schmittetal2001,Masonetal2012}.
Similarly, optical-/UV-/X-ray selected flux limited samples, are all
likely to have intrinsically more luminous Seyfert~2 galaxies
than Seyfert~1 galaxies \citep{Heckmanetal2005,WMW2009}.
In short, issues pertaining to sample selection are key
and new efforts to test the unified scheme hypothesis with superior
and well-chosen samples continue to be made
\citep[][and see also Section~\ref{sample}]{Laletal2011,Urry}.

This paper tests the validity and limitations of the unified scheme
hypothesis for Seyfert galaxies.
Here, we present results for a Seyfert galaxy sample
defined by \citet[][hereafter GMW09]{GMW2009},
using Very Large Array (VLA) survey images.
This large sample contains a diverse types of Seyfert galaxies.
Below, we first explain classifications of Seyfert galaxies
(Section~\ref{kinds}) and
give a brief description of the sample
(Section~\ref{sample}). We use the radio maps from
Faint Images of the Radio Sky at Twenty--Centimeters
\citep[FIRST;][]{becketetal} and
NRAO VLA Sky Survey \citep[NVSS;][]{Condonetal1998}
for the sample objects (Section~\ref{data}), and interpret our
results (Section~\ref{result}) and their
implications on the unified scheme hypothesis (Section~\ref{discuss}).
Finally, we summarize our conclusions in Section~\ref{summary}.

Following GMW09, we also adopt the same cosmological parameters
from \textit{Wilkinson Microwave Anisotropy Probe}.  Hence,
distance-dependent quantities are calculated assuming $H_0$ = 71
km~s$^{-1}$~Mpc$^{-1}$, $\Omega_{\rm m}$ = 0.27, and $\Omega_\Lambda$ = 0.73
\citep{Spergeletal2003}.
When archive radio data is available at other frequencies, we determine
radio spectral index, $\alpha$ which we define in the sense that
$S_\nu$ $\propto$ $\nu^\alpha$, where $S_\nu$ is the flux density
and $\nu$ is the frequency.
All coordinates mentioned below are for J2000 epoch.

\section{Background}

\subsection{Classifications of Seyfert Galaxies} \label{kinds}

In addition to Seyfert galaxies of type~1 and type~2;
\citet{Osterbrock81} and \citet{OP85}
introduced additional fractional classifications;
these appear to have a mix of properties from both types,
as the broad component
of H$\beta$ becomes weaker as compared to the narrower component,
the Seyfert type changes from 1.2 $\rightarrow$ 1.5 $\rightarrow$ 1.8 $\rightarrow$ 1.9.
Seyfert~1.2 and 1.5 galaxies have composite spectra,
with both broad and narrow components easily recognizable
and the former have broad component being stronger than narrow component,
whereas the latter have both components that are comparable.
In Seyfert~1.8 and 1.9 galaxies, in addition to the
narrow components, very weak but recognizable broad components of
H$\beta$ and H$\alpha$, or of H$\alpha$ alone, respectively are present
\citep{Osterbrock81}.
In terms of the unified scheme hypothesis
these intermediate Seyfert types are thought to lie at a varying angles between
pole-on and edge-on views,
with Seyfert 1.2 galaxies being close to pole-on, whereas Seyfert 1.9 galaxies
being close to edge-on.
\citet{OP85} proposed yet another type of Seyfert galaxies,
the narrow-line Seyfert~1 (NLS1) galaxies \citep{Valencia2012}.
These objects have permitted line-widths much smaller than typical
Seyfert~1 galaxies.  However, they differ from Seyfert~2 galaxies, in
the sense that optical spectra of NLS1 galaxies show several characteristics
normally associated with Seyfert~1 galaxies, such as [O\,III]/H$\beta$ ratios
of less than 3, permitted lines are broader than forbidden lines,
and blends of lines such as Fe\,II or [Fe\, VII] or [Fe\,X].
Again, in the unified scheme frame-work,
NLS1 galaxies could be sources with views very close to pole-on
\citep{Urry}.

\subsection{FHIL-Emitting Seyfert Galaxy Sample} \label{sample}

Our goal is to study the radio properties of Seyfert galaxies
and their implications on the unified scheme hypothesis, and
we use Seyfert galaxies listed in GMW09, with strong forbidden high-ionization line
(FHIL) emission.
This sample contains a diverse types of Seyfert galaxies,
including NLS1 galaxies.
It has
\begin{itemize}
\item[--] 12 NLS1,
\item[--] 14 type~1.0,
\item[--] 16 type 1.5,
\item[--] 2 type 1.9, and
\item[--] 18 type~2.0 Seyfert galaxies. 
\end{itemize}
Almost all known nearby, $z$ $\lesssim$ 0.1 Seyfert galaxy samples, e.g.,
the bright Seyfert galaxies sample \citep{giuricinetal},
the CfA Seyfert sample \citep{HB92},
the 12~{$\mu$}m Seyfert sample \citep{RMS1993,HuntMalkan99},
the far-infrared selected Seyfert galaxy sample \citep{royetal},
the matched Seyfert sample \citep{Laletal2011}, etc.
rarely have NLS1 galaxies in them and do not have appropriate proportion of Seyfert types
discussed above.
Other FHIL-emitting samples of Seyfert galaxies in the literature
\citep[e.g.,][]{DO1984,EAW97,MT1998,Veilleux1998,NTM2000}
are all relatively small and heterogeneous.
Instead, the GMW09 sample of Seyfert galaxies with strong FHILs is
one of the largest containing nearly equal proportion of diverse Seyfert types
and hence, by far the most homogeneous to date.
In order to discuss differences between NLS1, Seyfert~1 and Seyfert~2 galaxies
and its implications on the unified scheme hypothesis,
we henceforth address Seyfert~1.0 and 1.5, and Seyfert~1.9 and~2.0
as Seyfert~1, and Seyfert~2 galaxies, respectively.
Note that GMW09 found one galaxy with an unusual spectrum
that is not Seyfert-like (object 45, spectral type
``gal'' in GMW09), which was neither included by GMW09
nor it is included in the rest of this paper.
We also do not include radio galaxy 3C\,234 (object 22, spectral
type ``Seyfert~1.9''), though it was included by GMW09,
because it is a radio-loud active galactic nucleus \citep[AGN;][]{LRL1983}.
One another source (object 53, spectral type ``NLS1") is not covered by the FIRST survey
and is not detected in NVSS image;
this too is excluded.

\begin{table*}
\centering
\caption{Map parameters for the extended sources from the FHIL-emitting Seyfert galaxy sample}
\begin{tabular}{cccclr}
\toprule
{Object} & {Restoring} & {peak} & {r.m.s.} & {Contour levels} & {l.a.s.} \\
       & {beam} & \multicolumn{2}{c}{(mJy~beam$^{-1}$)} & ($\times$ {r.m.s.}) & \\
\midrule
SDSS\,J082930.59$+$081238.1 & 6.4\asec$\times$5.4\asec\ & 2.624 & 0.11 & $-$3, 3, 4, 6, 8, 10 & 6\asec \\
SDSS\,J092343.00$+$225432.6 & 5.4\asec$\times$5.4\asec\ & 5.516 & 0.08 & $-$3, 3, 4, 8, 12, 16, 20, 24, 32, 40, 48, 56 & 17\asec \\
SDSS\,J110704.52$+$320630.0 & 5.4\asec$\times$5.4\asec\ & 2.666 & 0.10 & $-$3, 3, 4, 8, 12, 16, 20, 24 & 6\asec \\
SDSS\,J115704.84$+$524903.7 & 5.4\asec$\times$5.4\asec\ & 2.142 & 0.10 & $-$3, 3, 4, 6, 8, 10 & 12\asec \\
SDSS\,J122930.41$+$384620.7 & 5.4\asec$\times$5.4\asec\ & 1.349 & 0.10 & $-$3, 3, 4, 6, 8, 10, 12 & 8\asec \\
SDSS\,J134607.71$+$332210.8 & 5.4\asec$\times$5.4\asec\ & 0.785 & 0.10 & $-$3, 3, 4, 6 & $\simeq$ 5\asec \\
SDSS\,J153552.40$+$575409.5 & 5.4\asec$\times$5.4\asec\ & 4.620 & 0.09 & $-$3, 3, 4, 8, 10, 12, 16, 20, 24, 32, 40, 48 & 11\asec \\
SDSS\,J220233.85$-$073225.0 & 6.4\asec$\times$5.4\asec\ & 1.296 & 0.09 & $-$3, 3, 4, 8, 10, 12 & 18\asec \\
\bottomrule
\end{tabular}
\label{map-param}
\end{table*}

Our final sample consists of 61 (from parent list of 64) Seyfert galaxies.
All these 61 objects have measured [Fe\,X], [Fe\,VII] and [O\,III] fluxes and
among these 59\% (36/61) objects have measured [Fe\,XI] fluxes,
which is twice as many as compared to earlier samples in the literature.
The mean redshifts and their dispersions for individual sub-classes of Seyfert galaxies are,
\begin{itemize}
\item[--] NLS1: 0.0764 $\pm$ 0.0313,
\item[--] Seyfert~1.0: 0.0962 $\pm$ 0.0612,
\item[--] Seyfert~1.5: 0.1061 $\pm$ 0.0701,
\item[--] Seyfert~1.9: 0.0877 $\pm$ 0.0341, and
\item[--] Seyfert~2.0: 0.0694 $\pm$ 0.0382.
\end{itemize}
These means are based on redshifts defined from the observed
wavelengths of the [S\,II] doublet.
It seems that the Seyfert 1.0 and 1.5 galaxies in the sample
are farther set of objects with respect to the
mean redshift for the sample $0.0869\pm0.0532$ possibly due to the bias mentioned above.
However, much of it is due to three outlier objects,
ID-12 (spectral type 1.0), and ID-16 and ID-28 (spectral type 1.5),
which are only objects with $z >$ 0.2 in the sample; and
excluding these objects, the means are $0.0840\pm0.0426$ and
$0.0843\pm0.0391$ for Seyfert~1.0 and Seyfert~1.5, respectively.
Barring this, the mean redshifts and their distributions
(see Figure~4, GMW09) for diverse types of Seyfert galaxies
agree among themselves.
In addition, the median redshifts of all Seyfert types also agree.
Therefore, apart from only possible bias,
bias against Seyfert~1 galaxies with the
broadest permitted lines and any FHIL-emitting sources dominated by
lines with lower ionization potentials,
presence of which is unclear (GMW09),
these Seyfert sub-classes do not differ in redshift distribution.
We, thus use this sample containing diverse types of Seyfert galaxies with FHIL features
to study the radio properties
and its implications on the unified scheme hypothesis.

\section{Data} \label{data}

Our earlier effort of investigating radio properties of type~2 SDSS quasars
\citep{LalHo2010} yielded a
detection rate of 59\% (35/59) at 8.4 GHz (X-band) survey.
This detection rate was essentially identical to that obtained from
FIRST survey images at 1.4 GHz with {5\asec}~resolution, which observed 56 out of the
59 sources in our sample and detected 35 (63\%),
even though the sensitivity of FIRST survey is nearly an order of magnitude
lower than that of our X-band survey (Lal \& Ho 2010).
Furthermore, the sensitivity of NVSS survey images
at 1.4 GHz with {45\asec}~resolution is $\sim$0.45 mJy~beam$^{-1}$
\citep{Condonetal1998},
a factor of three lower than the FIRST survey images and
surprisingly the detection rate was again identical.
Since, the sample of 61 Seyfert galaxies with FHIL-emitting features also come from
SDSS, we believe that deeper new observations may not be necessary
to understand basic radio properties.
We therefore use FIRST survey images (at 1.4\,GHz, $\sim$5$^{\prime\prime}$
resolution) together with NVSS images (at 1.4\,GHz, $\sim$45$^{\prime\prime}$ resolution),
and archive VLA data, when available
along with published optical and X-ray (GMW09) data
extensively to draw our conclusions.

\section{Results} \label{result}

\subsection{Maps and Source Parameters}

\begin{figure*}[t!]
\centering
\begin{tabular}{lll}
\includegraphics[width=5.5cm]{fig1a.ps} &
\includegraphics[width=5.5cm]{fig1b.ps} &
\includegraphics[width=5.5cm]{fig1c.ps} \\
\includegraphics[width=5.5cm]{fig1d.ps} &
\includegraphics[width=5.5cm]{fig1e.ps} &
\includegraphics[width=5.5cm]{fig1f.ps} \\
\includegraphics[width=5.5cm]{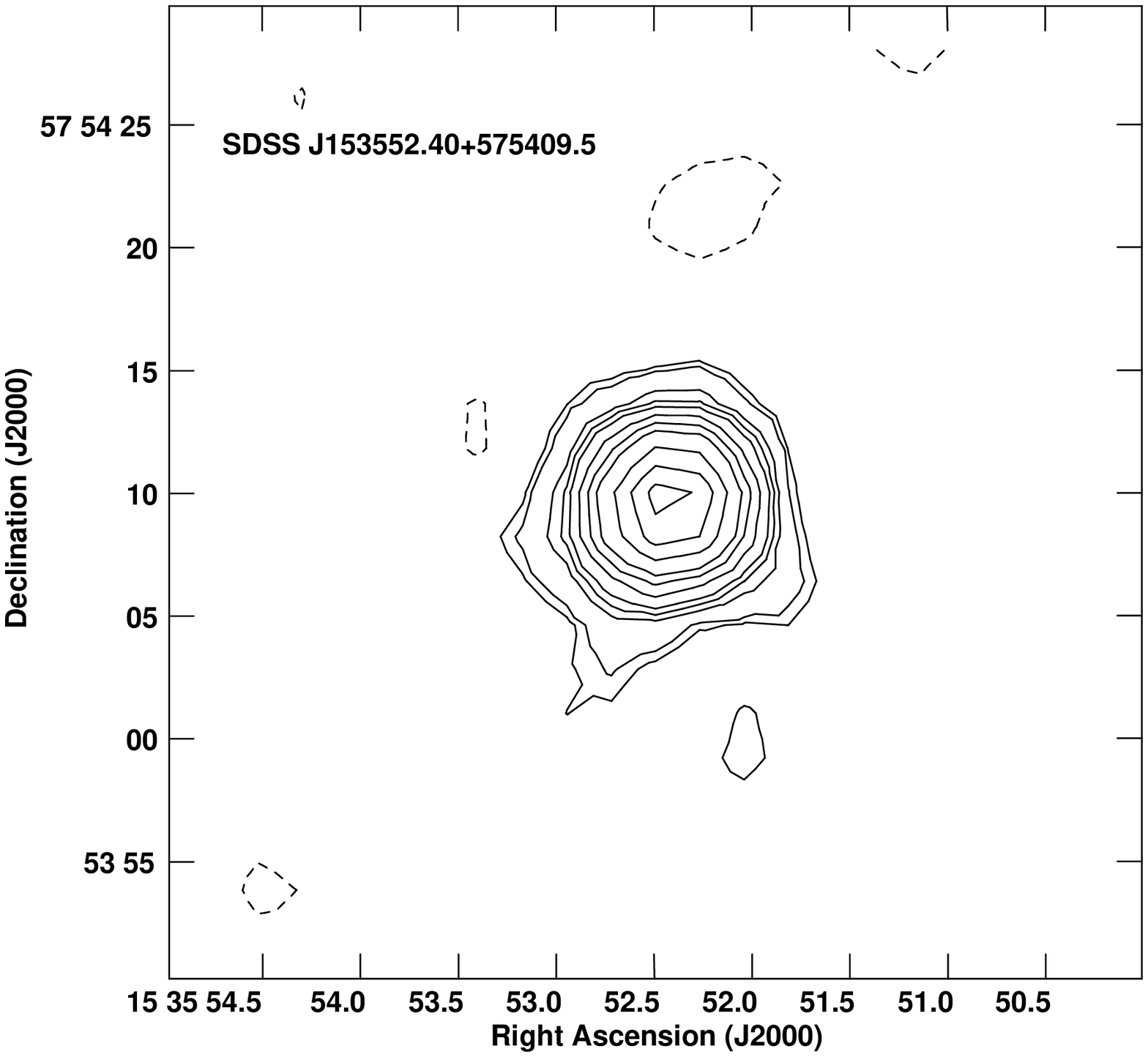} &
\includegraphics[width=5.5cm]{fig1h.ps} &
\end{tabular}
\caption{VLA $B$-array configuration, 1.4 GHz images with {5\asec}~resolution \citep[FIRST:][]{becketetal}
of selected FHIL-emitting Seyfert galaxies from SDSS showing extended radio structures.
The fields are centered on the optical positions given by GMW09.  The contour
levels and surface brightness peaks are listed in Table\ref{map-param}.}
\label{rad-images}
\end{figure*}

Galaxies were considered detected if the peak flux density
$S_{\rm Peak}^{\rm 1.4\,GHz}$ $>$ 5 $\times$ {r.m.s.} noise,
where the noise of each map was determined from a source--free region.
In total 30 sources from the sample of 61 were detected,
of which eight were resolved.
The radio images of eight sources showing extended, resolved structures,
are shown in Figure~\ref{rad-images}.
These source images are arranged in the order of increasing R.A., and
their restoring beams (with position angles = 0${^\circ}$),
root-mean-squared ({r.m.s.}) values and
contour levels in the maps are given in Table~\ref{map-param}.
The ratio of the peak surface brightness and the {r.m.s.} noise,
or the dynamic range for the FIRST survey, and the NVSS images
are between $\sim$5 and 66, and $\sim$6 and 43, respectively.
For the undetected sources, the flux density or its upper limit is equal to 5 $\times$ {r.m.s.}
The sizes (major and minor axes) and elongation direction (position angle)
of the detected components/sources were determined by fitting two-dimensional
Gaussians to each detected source and deconvolving those
Gaussians from the synthesized beams using Astronomical Image Processing
System (AIPS\footnote{\url{http://www.aips.nrao.edu}}) task JMFIT.
The data for the sources with complex morphologies, the
flux densities were found from integration over the maps covering
the source using AIPS task TVSTAT.

A summary of derived radio parameters from the FIRST survey images,
along with optical parameters is given in Table~\ref{rad_sum},
together with Seyfert galaxy types in the following order:
NLS1, Seyfert 1.0, 1.5, 1.9, and 2.0,
which are further arranged in the order of increasing R.A.
The individual columns are:
(1) Object name;
(2) ID Number as defined by GMW09; (3) spectral classification;
(4) redshift from the observed wavelength of the [S\,II] doublet;
(5) luminosity distance\footnote{computed using Ned Wright's online cosmology
calculator \url{http://www.astro.ucla.edu/$\sim$wright/CosmoCalc.html}};
(6) {r.m.s.} noise in the map;
(7) peak flux density;
(8) integrated flux density;
(9) deconvolved (half maximum) size of the source (major and minor axes);
(10) largest linear size ({l.l.s.});
(11) comments on the source structure; and
(12) integrated flux density as determined from the NVSS image.

For each object in Table~\ref{rad_sum},
we have classified the radio structure into following
categories \citep{UlvestadWilson1984,UlvestadHo2001}:
``U'' (unresolved), ``S'' (slightly resolved), 
``L'' (linear, usually core--jet or double),
and ``E'' (extended); where
slightly resolved and extended sources are those
whose deconvolved size is $\ge0.5$ times the synthesized beam width
\citep{UlvestadHo2001,LalHo2010}.
Whenever we obtain zero arcsec as the deconvolved size in any one
dimension, we ascribe ``U'' (unresolved) status to the object.

\subsection{Notes on Extended Sources}

There are 8 out of the 30 confirmed detections which are extended. \\ [-0.4cm]

{\sl SDSS\,J082930.59$+$081238.1} ---
This source appears extended in the FIRST image and
is classified as diffuse with faint extended emission.  The low-surface
brightness feature toward the northeast seen at 1.4 GHz is possibly a jet. \\ [-0.4cm]

%
{\sl SDSS\,J092343.00$+$225432.6} ---
The radio core is unresolved but the source appears to have
east--west elongation.  The low-surface brightness contours
suggest the source to be amorphous and 
is classified as an extended morphology. \\ [-0.4cm]

{\sl SDSS\,J110704.52$+$320630.0} ---
A marginal case of core-jet \citep{kharbetal} morphology.
The faint jet is seen as emanating from the core along the northern direction. \\ [-0.4cm]

{\sl SDSS\,J115704.84$+$524903.7} ---
The source is resolved, with typical core--jet morphology.

{\sl SDSS\,J122930.41$+$384620.7} ---
The source is marginally resolved with an extension along east. \\ [-0.4cm]

{\sl SDSS\,J134607.71$+$332210.8} ---
The low-surface brightness contours at 3$\sigma$--5$\sigma$ levels suggest that
the source is possibly a diffuse, with extended structure.
Assuming the structure at the phase-center and the extended feature
adjacent to it on the east to be real, the source is likely to be
of core--jet morphology. \\ [-0.4cm]

{\sl SDSS\,J153552.40$+$575409.5} ---
This object is also classified as Mrk\,290 (PG1534$+$580) at $z$ = 0.0296
(NED) in the literature.
The radio core is slightly resolved at 1.4 GHz.
An image using archive VLA $A$-array configuration data at 5.0\,GHz
(project-code AF0402, observe-date 2003-Jun-17)
shows the source to be unresolved.
The integrated spectral index using FIRST survey image and this archive data,
$\alpha_{\rm 1.4\,GHz}^{\rm 5.0\,GHz} = -0.73\pm0.04$, typical of many other
Seyfert galaxies \citep{Laletal2011}. \\ [-0.4cm]

%
{\sl SDSS\,J220233.85$-$073225.0} ---
This source has a double-lobed extended morphology at 1.4 GHz.
It is the southeast radio lobe, which is associated with the optical position.
Hence, it is possible that radio core is slightly resolved
showing a core--jet morphology.

\begin{table*}[p]
\centering
\caption{Summary of radio properties derived from FIRST survey images together with ID number (GMW09), Seyfert type, redshift and luminosity distance.}
\resizebox{175mm}{!}{
\begin{tabular}{lrccrlrrcccr}
\toprule
\multicolumn{1}{c}{Object} & ID & Class&$z_{[\rm S\,II]}$ & \multicolumn{1}{c}{$D_L$} & {r.m.s.} & $S_{\rm 1.4\,GHz}^{\rm Peak}$ & $S_{\rm 1.4\,GHz}^{\rm Int.}$ & $\theta_{\rm maj}$ $\times$ $\theta_{\rm min}$ & {l.l.s.} & Notes & $S_{\rm 1.4\,GHz}$ \\
       &    &      &        & (Mpc)   & \multicolumn{2}{c}{(mJy~beam$^{-1}$)} & (mJy)    & (arcsec$^2$) & (kpc) & & (mJy) \\
\multicolumn{1}{c}{(1)} & (2) & (3) & (4) & (5) & (6) & \multicolumn{1}{c}{(7)} & \multicolumn{1}{c}{(8)} & (9) & (10) & (11) & \multicolumn{1}{c}{(12)} \\
\midrule
SDSS\,J001852.47$+$010758.5 &  2 & NLS1 & 0.0583 &  257  & 0.142  &        & $<$0.71  &                       && &$<$2.25 \\
SDSS\,J092343.00$+$225432.6 & 20 & NLS1 & 0.0330 &  143  & 0.126  &  5.15  &    9.29  & 6.71  $\times$ 2.87   &11.1&L$+$E&   9.55 \\
SDSS\,J102235.15$+$022930.5 & 24 & NLS1 & 0.0701 &  312  & 0.138  &        & $<$0.69  &                       && &$<$2.25 \\
SDSS\,J103438.60$+$393828.3 & 25 & NLS1 & 0.0435 &  190  & 0.154  &  3.98  &    5.94  & 1.65  $\times$ 1.43   &&U&  24.19 \\
SDSS\,J110243.20$+$385152.6 & 27 & NLS1 & 0.1186 &  546  & 0.152  &        & $<$0.76  &                       && &$<$2.25 \\
SDSS\,J120932.94$+$322429.3 & 40 & NLS1 & 0.1303 &  605  & 0.143  &        & $<$0.72  &                       && &$<$2.25 \\
SDSS\,J131135.66$+$142447.2 & 46 & NLS1 & 0.1140 &  524  & 0.140  &        & $<$0.70  &                       && &$<$2.25 \\
SDSS\,J131957.07$+$523533.8 & 50 & NLS1 & 0.0922 &  417  & 0.138  &  3.05  &    2.72  &                       &&U&   3.15 \\
SDSS\,J161844.85$+$253907.7 & 59 & NLS1 & 0.0479 &  210  & 0.144  &        & $<$0.72  &                       && &$<$2.25 \\
SDSS\,J205822.14$-$065004.4 & 61 & NLS1 & 0.0740 &  331  & 0.139  &        & $<$0.70  &                       && &$<$2.25 \\
SDSS\,J220233.85$-$073225.0 & 62 & NLS1 & 0.0594 &  263  & 0.150  &  1.45  &    2.36  & 6.71  $\times$ 2.48   &20.4&L&$<$2.25 \\ [3pt]
SDSS\,J082930.59$+$081238.1 &  9 & S1.0 & 0.1295 &  601  & 0.144  &  1.51  &    2.32  & 4.37  $\times$ 3.50   &13.7&E&$<$2.25 \\
SDSS\,J083045.41$+$450235.9 & 11 & S1.0 & 0.1825 &  876  & 0.133  &        & $<$0.67  &                       && &$<$2.25 \\
SDSS\,J083658.91$+$442602.4 & 12 & S1.0 & 0.2544 & 1275  & 0.156  &  9.39  &   10.25  & 2.12  $\times$ 0.99   &&S&   6.08 \\
SDSS\,J084622.54$+$031322.2 & 14 & S1.0 & 0.1070 &  489  & 0.136  &        & $<$0.68  &                       && &$<$2.25 \\
SDSS\,J110756.55$+$474434.8 & 30 & S1.0 & 0.0727 &  324  & 0.135  &        & $<$0.68  &                       && &$<$2.25 \\
SDSS\,J112602.46$+$343448.2 & 33 & S1.0 & 0.0910 &  411  & 0.139  &        & $<$0.70  &                       && &$<$2.25 \\
SDSS\,J120422.15$-$012203.3 & 38 & S1.0 & 0.0834 &  375  & 0.148  &        & $<$0.74  &                       && &$<$2.25 \\
SDSS\,J121044.28$+$382010.3 & 41 & S1.0 & 0.0230 &   99  & 0.138  &  5.66  &    5.88  & 1.98  $\times$ 0.00   &&U&   7.14 \\
SDSS\,J131305.69$-$021039.3 & 47 & S1.0 & 0.0838 &  377  & 0.155  &        & $<$0.78  &                       && &$<$2.25 \\
SDSS\,J134607.71$+$332210.8 & 52 & S1.0 & 0.0838 &  377  & 0.130  &  1.11  &    1.41  & 5.81  $\times$ 0.50   &7.8&E&   2.52 \\
SDSS\,J143452.46$+$483942.8 & 54 & S1.0 & 0.0365 &  159  & 0.133  &        & $<$0.67  &                       && &$<$2.25 \\
SDSS\,J153552.40$+$575409.5 & 56 & S1.0 & 0.0304 &  131  & 0.146  &  5.11  &    5.32  & 1.15  $\times$ 1.05   &6.6&E&   4.31 \\
SDSS\,J161301.63$+$371714.9 & 58 & S1.0 & 0.0695 &  309  & 0.138  &        & $<$0.69  &                       && &$<$2.25 \\
SDSS\,J221542.30$-$003609.8 & 63 & S1.0 & 0.0994 &  452  & 0.135  &  1.55  &    1.62  & 2.33  $\times$ 0.00   &&U&$<$2.25 \\ [3pt]
SDSS\,J073126.69$+$452217.5 &  4 & S1.5 & 0.0921 &  417  & 0.128  &  2.75  &    2.68  & 1.67  $\times$ 0.00   &&U&   2.50 \\
SDSS\,J083045.37$+$340532.1 & 10 & S1.5 & 0.0624 &  276  & 0.141  &        & $<$0.71  &                       && &$<$2.25 \\
SDSS\,J085740.86$+$350321.7 & 16 & S1.5 & 0.2752 & 1395  & 0.141  &        & $<$0.71  &                       && &$<$2.25 \\
SDSS\,J091715.00$+$280828.2 & 18 & S1.5 & 0.1045 &  477  & 0.139  &        & $<$0.70  &                       && &$<$2.25 \\
SDSS\,J091825.79$+$005058.4 & 19 & S1.5 & 0.0871 &  393  & 0.145  &        & $<$0.73  &                       && &$<$2.25 \\
SDSS\,J094204.79$+$234106.9 & 21 & S1.5 & 0.0215 &   93  & 0.135  &  5.78  &    5.91  & 1.27  $\times$ 0.00   &&U&   5.52 \\
SDSS\,J101718.26$+$291434.1 & 23 & S1.5 & 0.0492 &  216  & 0.141  &        & $<$0.71  &                       && &$<$2.25 \\
SDSS\,J105519.54$+$402717.5 & 26 & S1.5 & 0.1201 &  554  & 0.132  &        & $<$0.66  &                       && &$<$2.25 \\
SDSS\,J110704.52$+$320630.0 & 28 & S1.5 & 0.2425 & 1207  & 0.145  &  3.03  &    2.82  & 1.86  $\times$ 0.80   &22.7&E&$<$2.25 \\
SDSS\,J110716.49$+$131829.5 & 29 & S1.5 & 0.1848 &  889  & 0.139  &        & $<$0.70  &                       && &$<$2.25 \\
SDSS\,J115226.30$+$151727.6 & 36 & S1.5 & 0.1126 &  517  & 0.146  &        & $<$0.73  &                       && &$<$2.25 \\
SDSS\,J122903.50$+$294646.1 & 42 & S1.5 & 0.0821 &  369  & 0.129  &        & $<$0.65  &                       && &$<$2.25 \\
SDSS\,J123149.08$+$390530.2 & 44 & S1.5 & 0.0683 &  304  & 0.150  &  2.19  &    2.31  & 3.31  $\times$ 0.00   &&U&   2.54 \\
SDSS\,J131348.96$+$365358.0 & 48 & S1.5 & 0.0670 &  298  & 0.143  &        & $<$0.72  &                       && &$<$2.25 \\
SDSS\,J163501.46$+$305412.1 & 60 & S1.5 & 0.0543 &  239  & 0.117  &  2.63  &    2.67  & 1.48  $\times$ 0.00   &&U&   3.53 \\
SDSS\,J235654.30$-$101605.5 & 64 & S1.5 & 0.0740 &  331  & 0.151  &  1.95  &    1.61  & 1.34  $\times$ 0.00   &&U&$<$2.25 \\ [3pt]
SDSS\,J001852.47$+$010758.5 &  1 & S1.9 & 0.0640 &  284  & 0.102  &        & $<$0.51  &                       && &$<$2.25 \\
SDSS\,J112602.46$+$343448.2 & 32 & S1.9 & 0.1114 &  511  & 0.145  &  3.64  &    3.81  & 2.56  $\times$ 0.00   &&U&   4.88 \\ [3pt]
SDSS\,J023301.24$+$002515.0 &  3 & S2.0 & 0.0224 &   96  & 0.095  &  3.89  &    3.98  & 1.68  $\times$ 0.00   &&U&   6.02 \\
SDSS\,J073638.86$+$435316.5 &  5 & S2.0 & 0.1140 &  524  & 0.135  &  4.59  &    4.26  & 0.54  $\times$ 0.00   &&U&   2.97 \\
SDSS\,J073650.08$+$391955.2 &  6 & S2.0 & 0.1163 &  535  & 0.131  &  8.13  &    8.65  & 1.75  $\times$ 0.86   &&S&   8.82 \\
SDSS\,J080707.18$+$361400.5 &  7 & S2.0 & 0.0324 &  140  & 0.140  &  0.71  &    0.71  & 0.54  $\times$ 0.50   &&U&   4.35 \\
SDSS\,J081153.16$+$414820.0 &  8 & S2.0 & 0.0999 &  454  & 0.131  &        & $<$0.66  &                       && &$<$2.25 \\
SDSS\,J084215.30$+$402533.3 & 13 & S2.0 & 0.0553 &  244  & 0.152  &  1.68  &    1.43  &                       &&U&$<$2.25 \\
SDSS\,J085332.22$+$210533.7 & 15 & S2.0 & 0.0719 &  321  & 0.137  &        & $<$0.69  &                       && &$<$2.25 \\
SDSS\,J085810.64$+$312136.3 & 17 & S2.0 & 0.1389 &  649  & 0.141  &  2.20  &    2.05  & 1.52  $\times$ 0.00   &&U&$<$2.10 \\
SDSS\,J110929.10$+$284129.2 & 31 & S2.0 & 0.0329 &  143  & 0.149  &        & $<$0.75  &                       && &$<$2.25 \\
SDSS\,J113917.17$+$283946.9 & 34 & S2.0 & 0.0234 &  101  & 0.153  &        & $<$0.77  &                       && &$<$2.25 \\
SDSS\,J114216.88$+$140359.7 & 35 & S2.0 & 0.0208 &   89  & 0.135  &  1.55  &    1.84  & 3.18  $\times$ 1.17   &&S&$<$2.25 \\
SDSS\,J115704.84$+$524903.7 & 37 & S2.0 & 0.0356 &  155  & 0.144  &  1.19  &    4.22  & 8.73  $\times$ 8.53   &8.4&S$+$L&   2.81 \\
SDSS\,J120735.06$-$001550.3 & 39 & S2.0 & 0.1104 &  506  & 0.138  &        & $<$0.69  &                       && &$<$2.25 \\
SDSS\,J122930.41$+$384620.7 & 43 & S2.0 & 0.1024 &  467  & 0.141  &  1.53  &    2.90  & 7.42  $\times$ 2.71   &14.9&E&   3.35 \\
SDSS\,J131639.75$+$445235.1 & 49 & S2.0 & 0.0911 &  412  & 0.135  &  4.23  &    4.53  & 2.05  $\times$ 0.36   &&S&   5.68 \\
SDSS\,J132346.00$+$610400.2 & 51 & S2.0 & 0.0715 &  319  & 0.151  &  9.16  &    9.01  & 1.17  $\times$ 0.00   &&U&  19.49 \\
SDSS\,J153222.32$+$233325.0 & 55 & S2.0 & 0.0465 &  204  & 0.162  &  1.10  &    0.98  & 2.42  $\times$ 0.00   &&U&$<$2.25 \\
SDSS\,J160948.21$+$043452.9 & 57 & S2.0 & 0.0643 &  285  & 0.149  &  4.27  &    4.75  & 2.41  $\times$ 0.74   &&S&   3.89 \\
\bottomrule
\end{tabular}
}
\label{rad_sum}
\end{table*}

\section{Discussion} \label{discuss}

\subsection{The Radio Properties}

\subsubsection{Radio Detection Rate} \label{det-rate}

Of the 61 FHIL-emitting Seyfert galaxies in the sample, 30 (49\%)
have been detected down to the FIRST survey {r.m.s.} sensitivity limit of
0.15 mJy beam$^{-1}$ \citep{becketetal}.
This detection rate further lowers, 34\% (21/61)
if we instead consider NVSS data.
Furthermore, this detection rate is:
\begin{enumerate}
\item{lower than the detection rate of type~2 SDSS quasars sample
\citep[][median redshift = 0.427]{LalHo2010},
$\sim$59\% (35/59) and 63\% (35/56)
using X-band images and using FIRST survey images, respectively;}
\item{lower than the detection rate of complete sample of local Seyfert galaxies
\citep[][median redshift = 0.003]{PanessaGiorletti2013}: 74\% (17 of 23, milliarcsec-scale resolution)
and 100\% (23 of 23, VLA, arcsec-scale resolutions);}
\item{lower than the detection rate of matched sample of Seyfert galaxies
\citep[][median redshift = 0.019]{Laletal2011}: 100\% (20 of 20; both, milliarcsec-scale and
arcsec-scale resolutions);}
\item{lower than the detection rate of extended 12~{$\mu$}m Seyfert galaxy sample
\citep[][median redshift = 0.008]{RMS1993,Theanetal00}: 86\%
(75 of 87; VLA $A$-array configuration, sub-arcsec-scale resolutions);}
\item{lower than the detection rate of CfA Seyfert galaxy sample
\citep[][median redshift = 0.0141]{HB92,Kukulaetal95}:
81\% (39 of 48; VLA $A$-array configuration, sub-arcsec-scale resolutions) and
88\% (42 of 48; VLA $C$-array configuration, arcsec-scale resolutions);}
\item{lower than the detection rate of bright Seyfert galaxies sample
\citep[][median redshift = 0.009]{giuricinetal}: 90\%;}
\item{(nearly) similar to the detection rate of far-infrared selected
Seyfert galaxy sample: $\sim$39\% \citep[][median redshift = 0.031]{royetal}.}
\end{enumerate}
Note that barring type~2 SDSS quasar sample \citep{LalHo2010},
all other Seyfert galaxy samples are nearby (redshift $\lesssim$ 0.03),
and except
radio observations of two Seyfert galaxy samples,
\citet{royetal} and \citep{PanessaGiorletti2013}, which are
at higher angular resolutions, the radio data of all other samples
are at similar, arcsecond-scale resolutions.
Among these nearby Seyfert galaxy samples,
the detection rates of FHIL-emitting Seyfert galaxy and
far-infrared selected samples are similar.
Looking more closely only at the detected sources,
the sample size of 21 objects; namely,
\begin{enumerate}
\item[--] three (out of 11) NLS1,
\item[--] four (out of 14) Seyfert~1.0,
\item[--] four (out of 16) Seyfert~1.5,
\item[--] one (out of 2) Seyfert~1.9 and
\item[--] nine (out of 18) Seyfert~2.0.
\end{enumerate}
Thus, the detection rates of NLS1, Seyfert 1 and Seyfert~2
galaxies are $0.25\pm0.16$, $0.27\pm0.11$, and $0.50\pm0.07$, respectively.
Alternatively, the radio detection rate of compact radio structure
for NLS1, Seyfert 1 and Seyfert~2 is consistent with the unified scheme hypothesis.

\begin{figure*}[t!]
\centering
\includegraphics[angle=270,width=8.5cm]{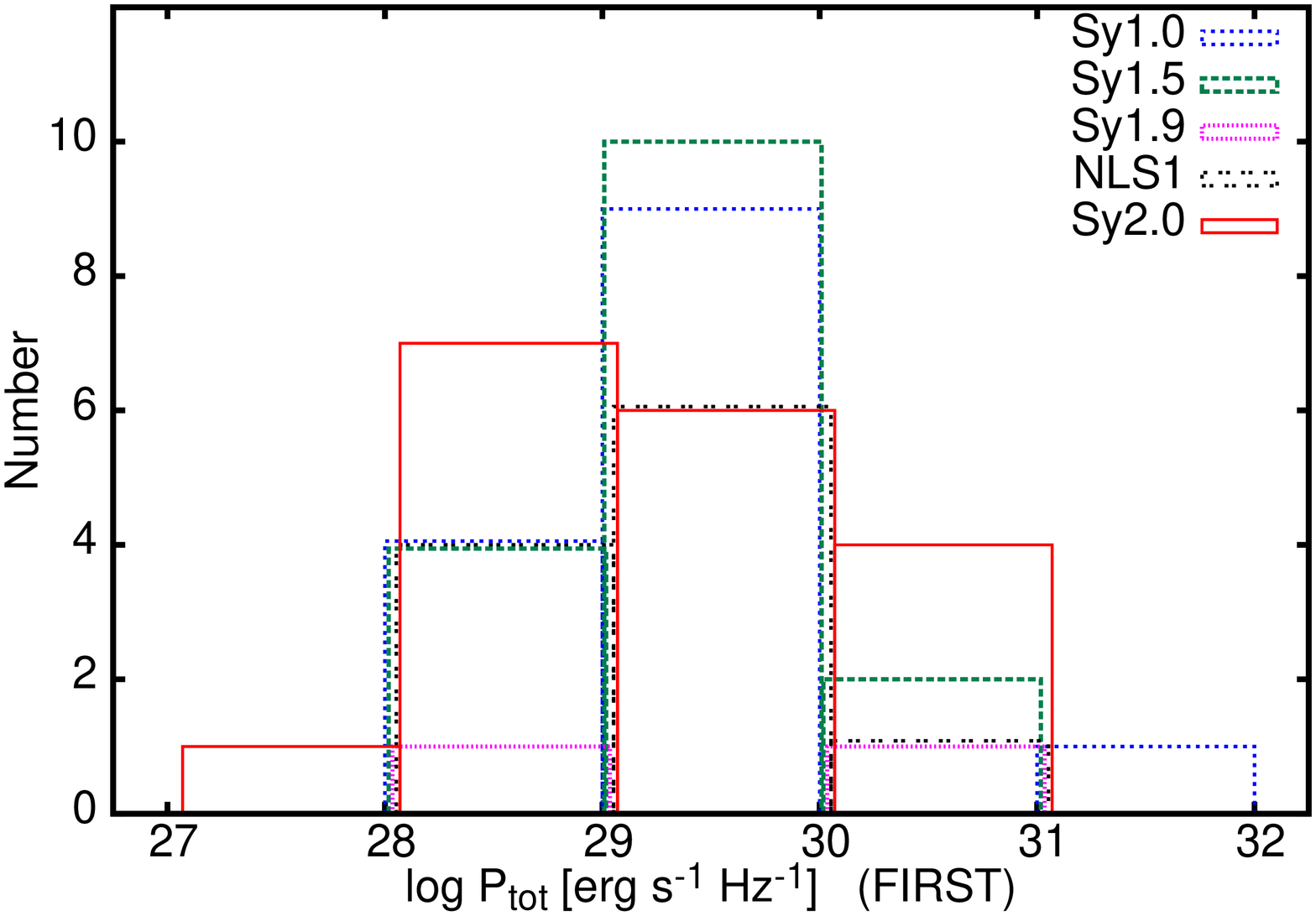}\hspace{3mm}
\includegraphics[angle=270,width=8.5cm]{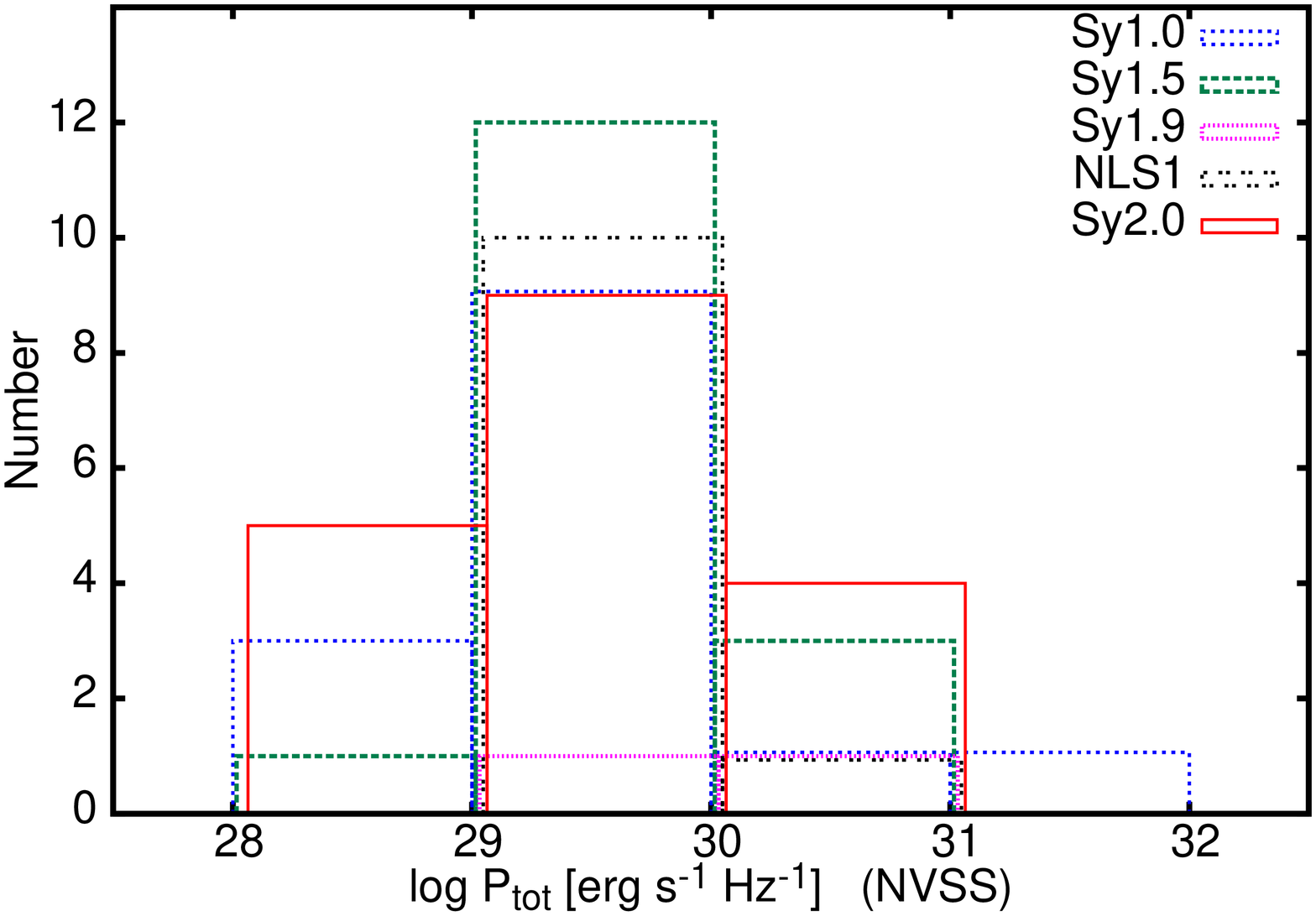}
\caption{Histograms showing distributions of total detected radio
luminosity (erg~s$^{-1}$~Hz$^{-1}$) for the FHIL-emitting Seyfert galaxies from
SDSS at 1.4 GHz using VLA $B$-array \citep[left panel, FIRST:][]{becketetal} and
the radio luminosity (erg~s$^{-1}$~Hz$^{-1}$) at 1.4 GHz using VLA
$D$-array \citep[right panel, NVSS:][]{Condonetal1998}.
The undetected sources are not explicitly denoted,
they all lie at the lower end of the luminosity, and
the bins widths are chosen large enough to incorporate upper limits.
Line colours and styles are:
NLS1 = black, Seyfert~1.0 = blue, Seyfert~1.5: dark-green, Seyfert~1.9 = magenta, and Seyfert~2.0 = red.
The overlapping lines are marginally shifted with respect to
each other for clarity.}
\label{first-nvss}
\end{figure*}

\citet{UWS81}, \citet{Whittleetal86}, \citet{Evansetal91} and \citet{FWS98}
noted that compact radio emission is closely
associated with individual narrow-line region clouds,
and \citet[][and see also Norris {et~al.} 1992]{royetal} first invoked a model that
free--free absorption by these clouds may explain
the low detection rate in FHIL-emitting Seyfert galaxies.
In particular, when observing Seyfert galaxies
showing very low radio emission from the nucleus,
either (i) the narrow-line region or
(ii) the individual narrow-line region clouds would absorb radio emission; i.e.,
if the optical depths are above unity due to free--free absorption then
either the narrow-line region or
the individual narrow-line clouds will block our view of the core and we would not
detect it in radio band.
Although this model involve optical depth effects in the narrow-line region,
its geometry and filling factor, it reconciles our low radio detection rate.
Furthermore, in this model, the narrow-line region clouds
would become optically thin at higher frequencies and
optically thick at further lower frequencies \citep{Norrisetal1992}.
So new radio observations at a sufficiently high frequencies should find higher
detection rates and similarly, at sufficiently low frequencies should find
lower detection rates, which is a possible test
if free--free absorption by these clouds
is indeed responsible for the low detection rate at 1.4\,GHz for the FHIL-emitting
Seyfert galaxies in the sample.

\subsubsection{Projected Linear Sizes}

Twenty-two of 30 ($\sim$73\%) detected objects
are classified as ``U'' or ``S'', unresolved or slightly resolved
and the radio emission in these is largely from the central active nucleus.
This result shows that a significant fraction of Seyfert galaxies
do not have double/triple radio sources, instead they have unresolved
structures on a scale of several hundred parsecs to a few kiloparsecs.
The double radio sources that have been found in Seyfert galaxies generally
have angular separations smaller than the resolution of the present data
\citep{Kukulaetal95,Theanetal00,Laletal2004}.
Although the exact number must await high-resolution radio maps
of the complete FHIL-emitting Seyfert sample,
eight (out of detected 30 = 27\%) sources (Figure~\ref{rad-images}) have
marginally extended, about a few kiloparsec scale to a few tens of
kiloparsec scale structures.
The largest projected linear sizes ({l.l.s.}),
determined from largest-angular-sizes ({l.a.s.}) listed in Table~\ref{map-param},
for eight extended sources shown in Figure~\ref{rad-images}
corresponding to the 3$\sigma$ contour level
of the 1.4\,GHz maps are listed in Table~\ref{rad_sum}.

Clearly Seyfert~1 and NLS1 galaxies have larger range in projected linear
sizes as compared to Seyfert~2 galaxies.
Unfortunately with only eight sources, three Seyfert~1, three NLS1 and
two Seyfert~2, this is small number statistics.
The statistical test\footnote{Since, we are dealing with small number
statistics, we use instead Mann--Whitney U test, a non-parametric
statistical hypothesis test for small sample sizes \citep{SiegelCastellan}.}
gives a poor significance level
of 0.25 that the distribution of projected linear sizes of
NLS1, Seyfert~1 and Seyfert~2 galaxies are same.
Higher-resolution observations are therefore needed to resolve the FIRST survey
images into structures with physical sizes less than a kiloparsec,
as often seen in nearby Seyfert galaxies
\citep[{e.g.,}][]{Kukulaetal95,Theanetal00,Laletal2004}.
These resolved morphologies would provide improved statistics and
test the distribution of projected linear sizes at a higher significance level.

\subsubsection{Kiloparsec-Scale Radio Luminosities}
\label{kpc-rad-lum}

Figure~\ref{first-nvss} shows the distribution of the radio luminosity
detected on 5$^{\prime\prime}$-scales (FIRST images) and
45$^{\prime\prime}$-scales (NVSS images) for all Seyfert sub-classes.
Radio luminosities for our sample sources are $K$-corrected,
rest-frame values determined
from observed flux densities at 1.4\,GHz and
corresponding luminosity distances \citep{LalHo2010}.
Since majority (22 of the 30 confirmed detections) of the sources are unresolved
and correspond to core emission, we assume
a typical spectral index, $\alpha$ = $+$0.5 from cores of Seyfert galaxies
\citep{KellermannOwen}.
We have also chosen the radio luminosity bin widths sufficiently larger than
the {r.m.s.} noise in the maps to account for
upper limits for undetected sources, which are not explicitly denoted in
Figure~\ref{first-nvss}.
Although non-detections lie in the luminosity-bins corresponding
to the low end of radio powers in both panels,
the distribution of radio powers at 1.4 GHz is continuous,
showing no separation between the detected and the non-detected objects.
Furthermore,
changing the radio luminosity bin size does not make the distribution
bimodal (or multimodal).
This suggests that the detected and undetected sources from the
FHIL-emitting Seyfert galaxy sample irrespective of Seyfert class
denote a single population as a whole.

In addition, the radio power distributions are not significant different between
NLS1 galaxies, Seyfert~1 galaxies and Seyfert~2 galaxies.
The K--S test gives a significance level
of 0.026 or better that the distribution of Seyfert~1 along with NLS1 galaxies
and Seyfert 2 galaxies are same for the FIRST and NVSS data.
Thus, we conclude that the radio power distributions
of Seyfert galaxy types are similar
and are consistent with the expectations of the unified scheme.

\subsubsection{Nature of Radio Emission}

\begin{figure}[t!]
\centering
\includegraphics[angle=270,width=8.35cm]{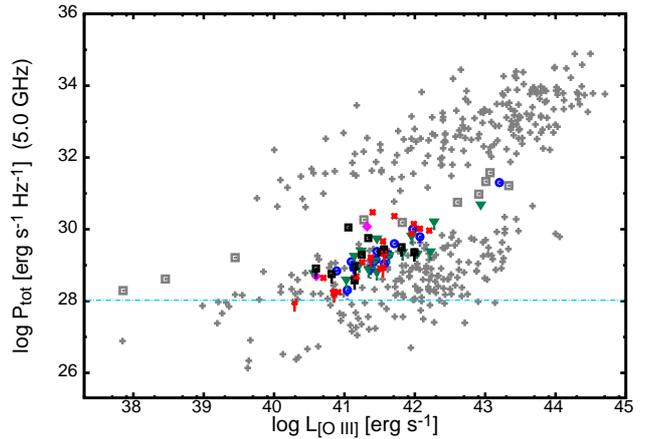}
\caption{Correlation between radio power and [O\,III]\,$\lambda$\,5007\,\AA\ luminosity.
The grey colour plus symbols come from the sample of \citet{XuLivioBaum1999},
with the grey colour squares representing their radio-intermediate sources.
The point colours and styles are:
NLS1 = black squares, Seyfert~1.0 = blue circles, Seyfert~1.5: dark-green triangles,
Seyfert~1.9 = magenta diamonds, and Seyfert~2.0 = red stars.
Upper limits are marked with downward-pointing lines.
The dashed-dotted line shows the expected radio luminosity at 5\,GHz
predicted from star formation rate (see Section~\ref{core-det}).
}
\label{xu-gel-samp}
\end{figure}

\citet{baumheckman} and \citet{RawlingsSaunders1991} showed that a
correlation exists between the radio luminosity at 5\,GHz
and the [O\,III]\,$\lambda$\,5007\,\AA\ narrow-line luminosity.
Figure~\ref{xu-gel-samp} illustrates the distribution of
integrated radio power (at 5 GHz) versus
[O\,III] luminosity for the large,
heterogeneous sample of AGNs compiled by \citet{XuLivioBaum1999},
after correcting for the cosmology.
The data for FHIL-emitting Seyfert galaxies come from 1.4 GHz;
and to be consistent with \citet{XuLivioBaum1999},
we converted these measurements to 5~GHz assuming a flat spectrum
spectral index of $\alpha$ = 0 and determine the monochromatic power.
These monochromatic powers change at the most by a factor of 12\%,
if instead we assume a slightly inverted or steep spectra
(see also Section~\ref{kpc-rad-lum}) and
it does not change our interpretations below.
Note that we have computed [O\,III] luminosities using
combined fluxes of [O\,III] core and [O\,III] wing components
(supplementary material: Table~A4, Column~8; GMW09).
It is clear that AGNs separate
into the two families of radio-loud and radio-quiet AGNs with a
significant gap between them; i.e., the
radio luminosities are different by a factor of 10$^3$--10$^4$
between the two groups at a given [O\,III] luminosity.
All the objects in FHIL-emitting Seyfert galaxy sample are of radio-quiet nature,
with $\sim$8\% that lie in between the two families of radio-loud and radio-quiet AGNs,
also called radio-intermediate AGNs.
It is not surprising that all the non-detected sources are radio-quiet.
Linear fits to the FHIL-emitting Seyfert galaxy sample excluding the
radio-intermediate objects yield
$$
\log L_{\rm 5~GHz} = (0.45\pm0.11) \log L_{\rm [O\,III]} + (5.6\pm0.9),
$$
which is consistent within error bars with the linear fit by
\citet[excluding the radio-intermediate objects;][]{XuLivioBaum1999} obtained for
the radio-quiet AGNs.

Furthermore, assuming 50\% of the flux density detected in the FIRST survey images
is also detected at milliarcsec-scales \citep{Laletal2011}, the inferred brightness
temperature, $T_{\rm B}$ $\geq$ 5 $\times$ 10$^8$~K for our sample objects, which is
again typical of Seyfert galaxies \citep{BF2011}.

\subsubsection{Relativistic Beaming} \label{beaming}

In radio galaxies, one-sided structures are often associated with
relativistic jets and the jet to counter-jet ratios are used to obtain
quantitative estimates of relativistic beaming.
Here, unfortunately a large fraction of sources are non-detected
($\sim$51\% in FIRST and $\sim$67\% in NVSS survey images) and
a majority of the rest of the detected sources are unresolved
(22 objects out of 30).
The FIRST survey images probes structures on scales smaller than
the NVSS survey images (see Section~\ref{data}).
Therefore, we assume any difference of
the emissions between NVSS and FIRST images
as extended emission, which is not Doppler boosted;
and instead, if beaming is present, the emission detected in FIRST image
would possibly be Doppler boosted \citep{Laletal2011}.
Hence, the ratio of the possibly boosted and the extended radio flux densities,
or the $R$-parameter
could be used here to investigate relativistic beaming. 
Since a majority of sources are not detected,
we focus only on the detected sources (see also Section~\ref{det-rate}). 
The ratio of flux density for these detected (20 of 61) sources in
both, FIRST and NVSS survey images is close to 1.0
with mean and median being $1.06\pm0.04$ and 1.02, respectively.
Clearly, much higher resolution, scales of milli-arcsecond would be required to
address relativistic beaming.
Additionally, the use of $R$-parameter as measure of relativistic beaming
comes with a caveat.
\citet{Laletal2011} have shown that Seyfert galaxies show radio variability
and the FIRST and the NVSS
data for the FHIL-emitting Seyfert sample are not simultaneous.
Therefore, here we have made an assumption that
these sample objects do not show radio variability \citep[see also][]{Mundeletal2009}.

Additionally, \citet{FSP96} showed that Lorentz factors of $\gamma$ = 2--4 in
radio-quiet AGNs are sufficient to boost the radio emission
into the radio-intermediate regime or into the radio-loud regime.
Five objects from FHIL-emitting Seyfert galaxy sample
are of radio-intermediate kind:
\begin{enumerate}
\item[--] one (out of 11) NLS1,
\item[--] one (out of 14) Seyfert~1.0,
\item[--] one (out of two) Seyfert~1.9 and
\item[--] two (out of 18) Seyfert~2.0.
\end{enumerate}
Although this is a small number statistics,
majority are non-type~1 Seyfert galaxies,
and it is well known that Seyfert~2 galaxies being edge-on counterparts
of beamed population, which are not expected to show relativistic beaming.
Therefore, it is possible that FHIL-emitting Seyfert galaxy sample are
radio-quiet and there is no evidence of relativistic beaming consistent
with the predictions of unified scheme hypothesis,
which is further supported by other Seyfert galaxy samples \citep{Laletal2011}.

\begin{figure*}[!t]
\centering
\includegraphics[angle=270,width=8.5cm]{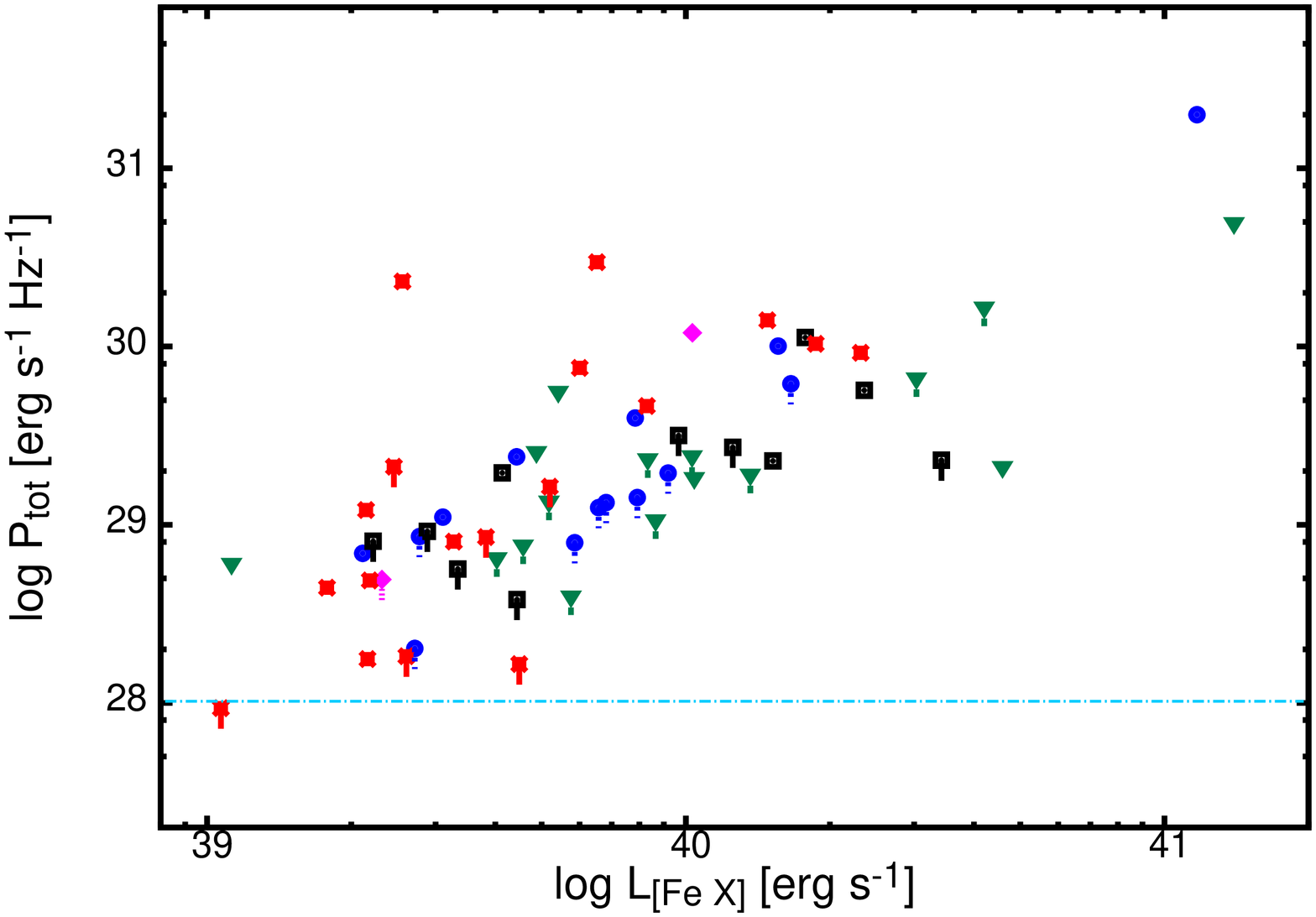} \hspace{2mm}
\includegraphics[angle=270,width=8.5cm]{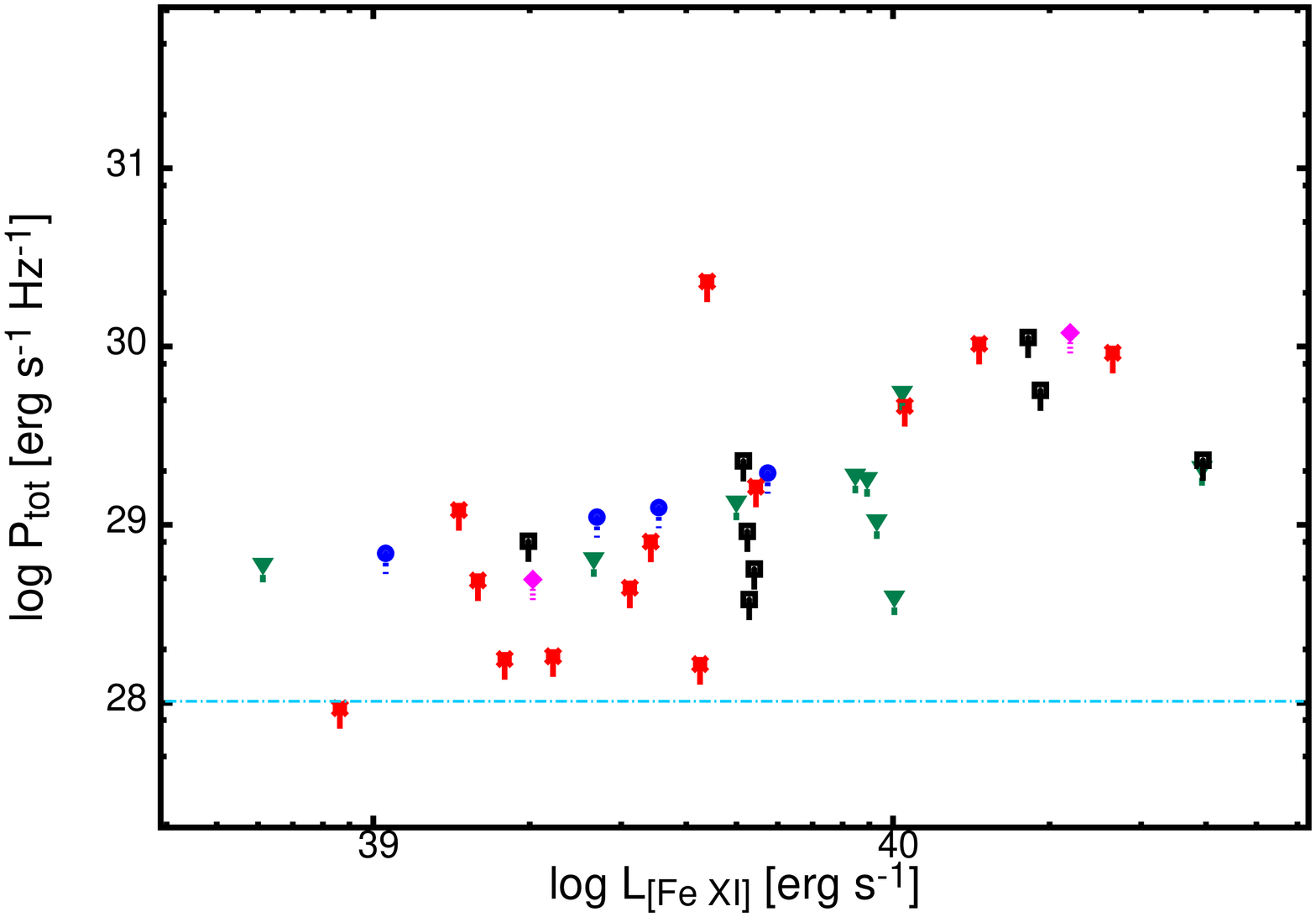} \\
\includegraphics[angle=270,width=8.5cm]{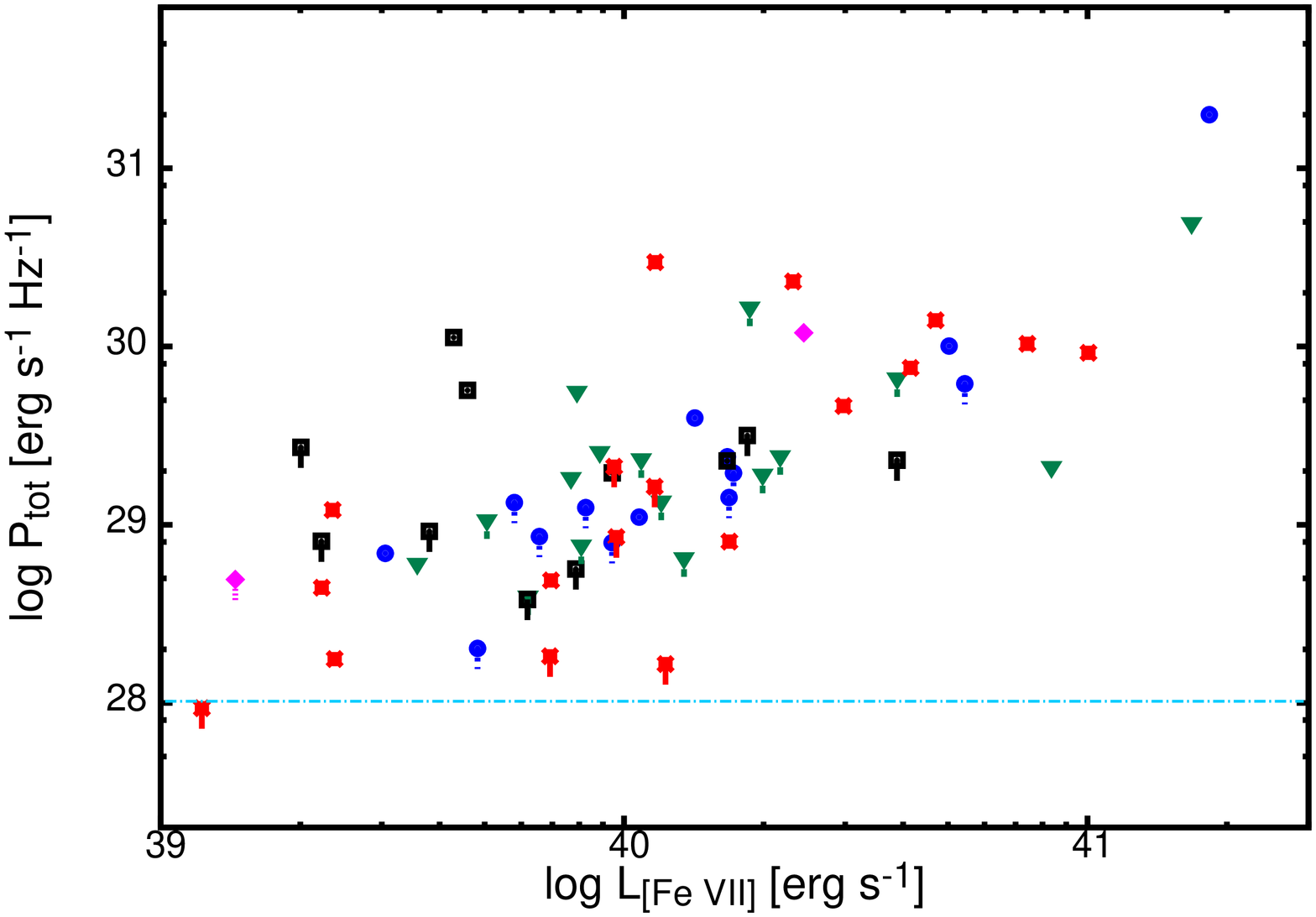} \hspace{2mm}
\includegraphics[angle=270,width=8.5cm]{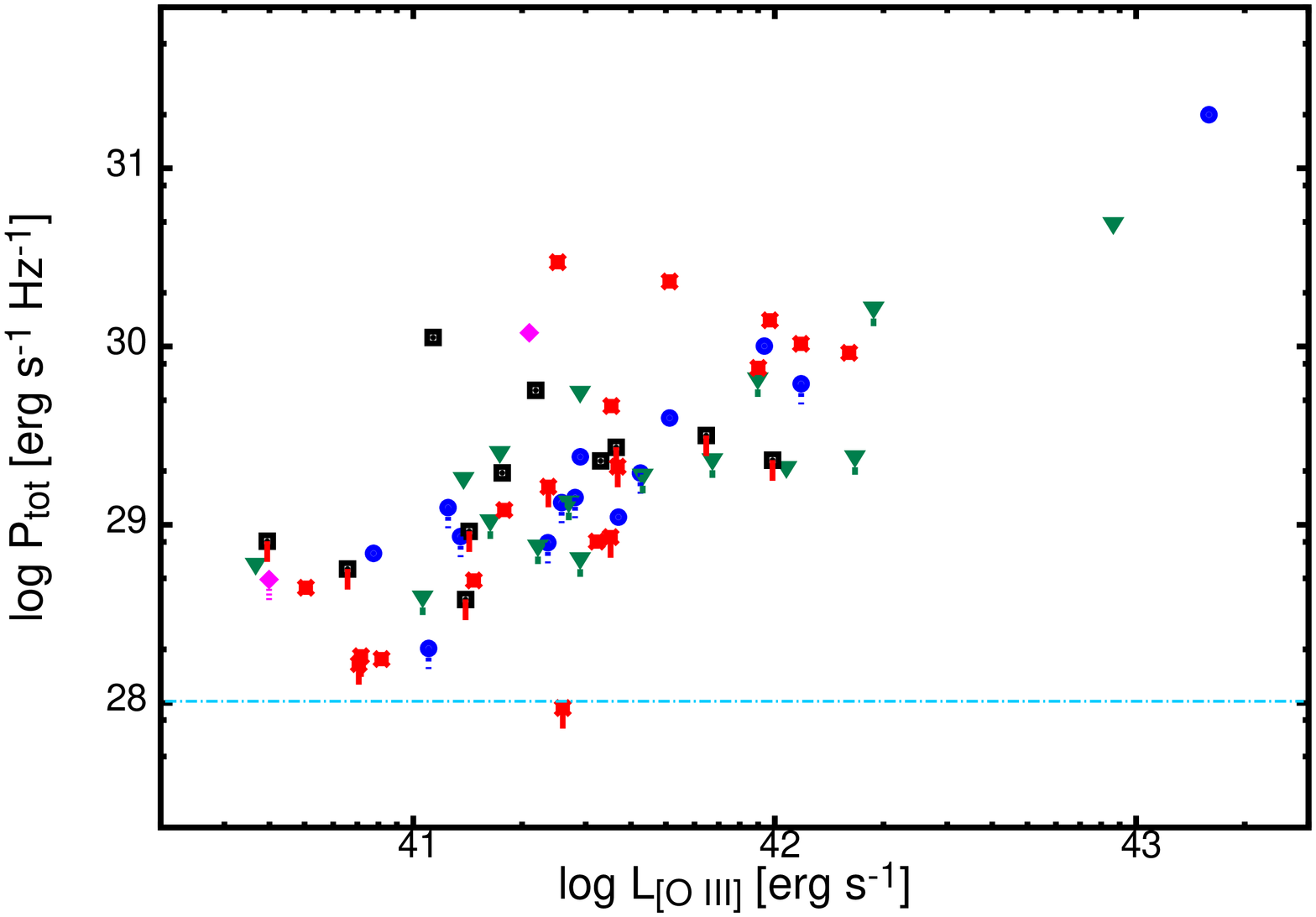} \\
\includegraphics[angle=270,width=8.5cm]{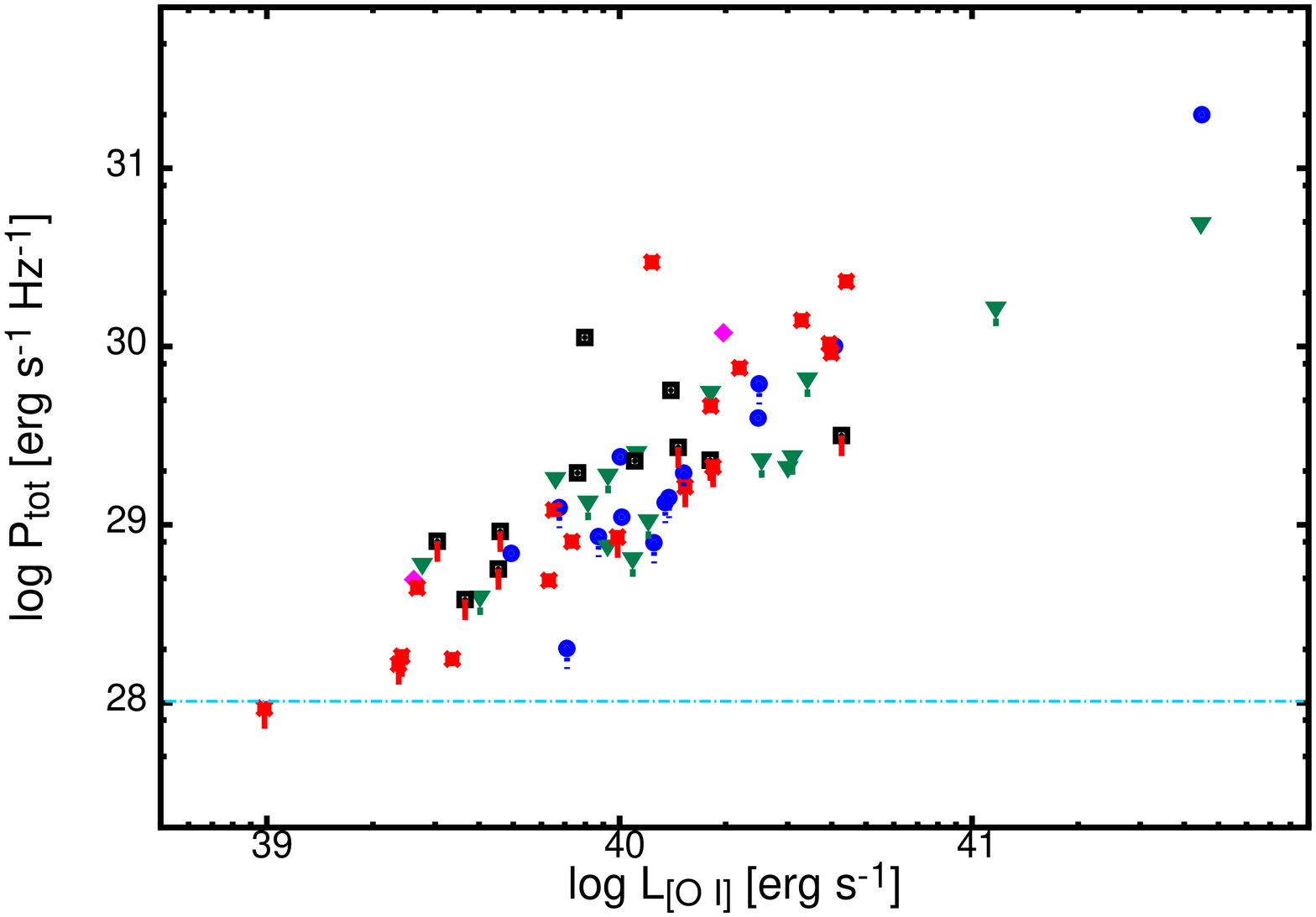} \hspace{2mm}
\includegraphics[angle=270,width=8.5cm]{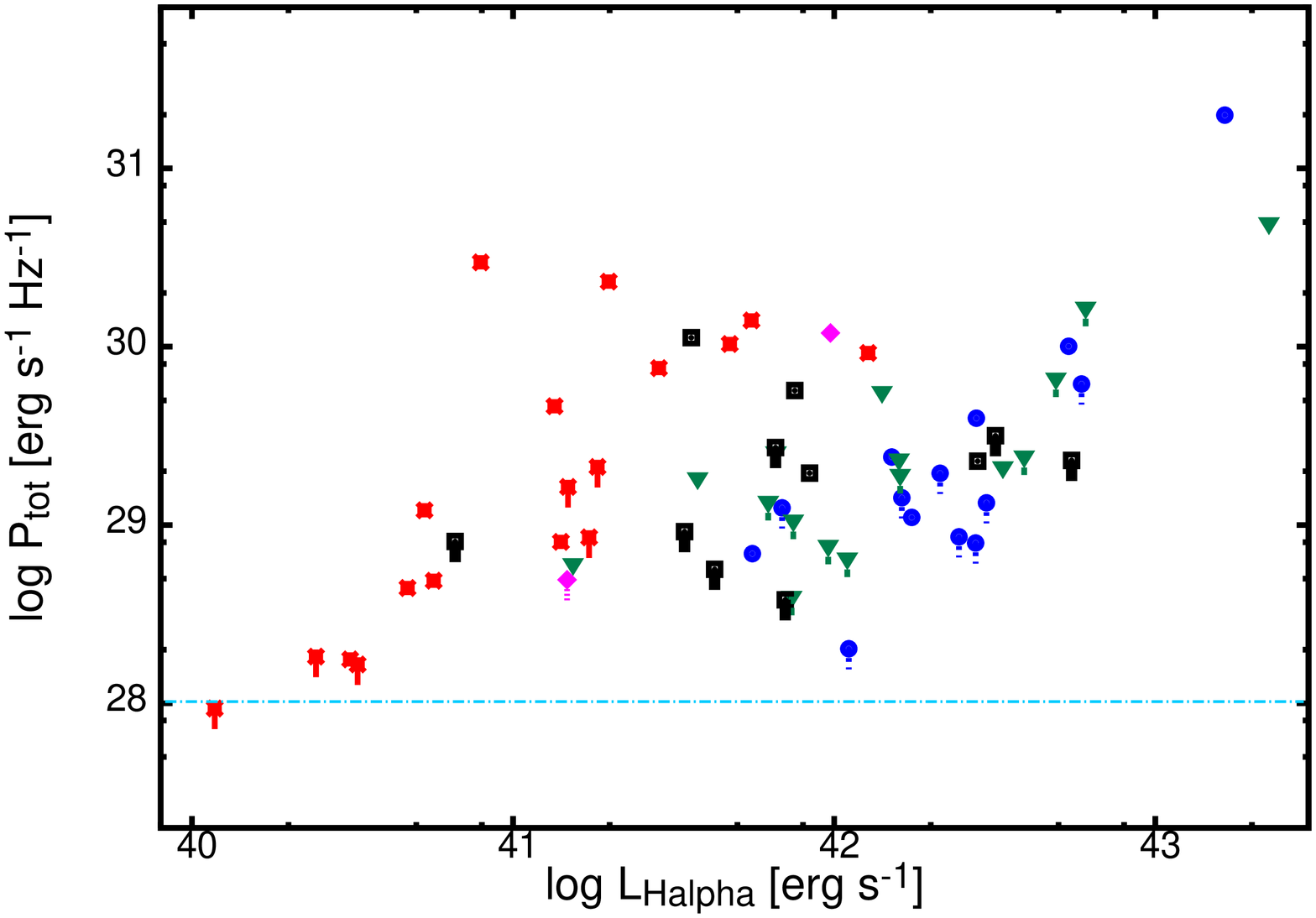}
\caption{Radio power at 1.4~GHz from VLA~$B$-array configuration
versus luminosities of
[Fe\ X], [Fe\ XI], [Fe\ VII], [O\,III], [O\,I], and H$_{{\alpha}}$.
The point colours and styles are:
NLS1 = black squares, Seyfert~1.0 = blue circles,
Seyfert~1.5: dark-green triangles,
Seyfert~1.9 = magenta diamonds, and Seyfert~2.0 = red stars.
Upper limits are marked with downward-pointing lines.
The dashed-dotted line shows the expected radio luminosity at 5\,GHz
predicted from star formation rate (see Section~\ref{core-det}).
Error bars (given in Tables A1--A7: supplementary material, GMW09, are not shown,
since they are typically $\sim$2 times the size of the downward-pointing lines.}
\label{fe-o-h-corr}
\end{figure*}

\begin{figure*}[t!]
\centering
\includegraphics[angle=270,width=8.5cm]{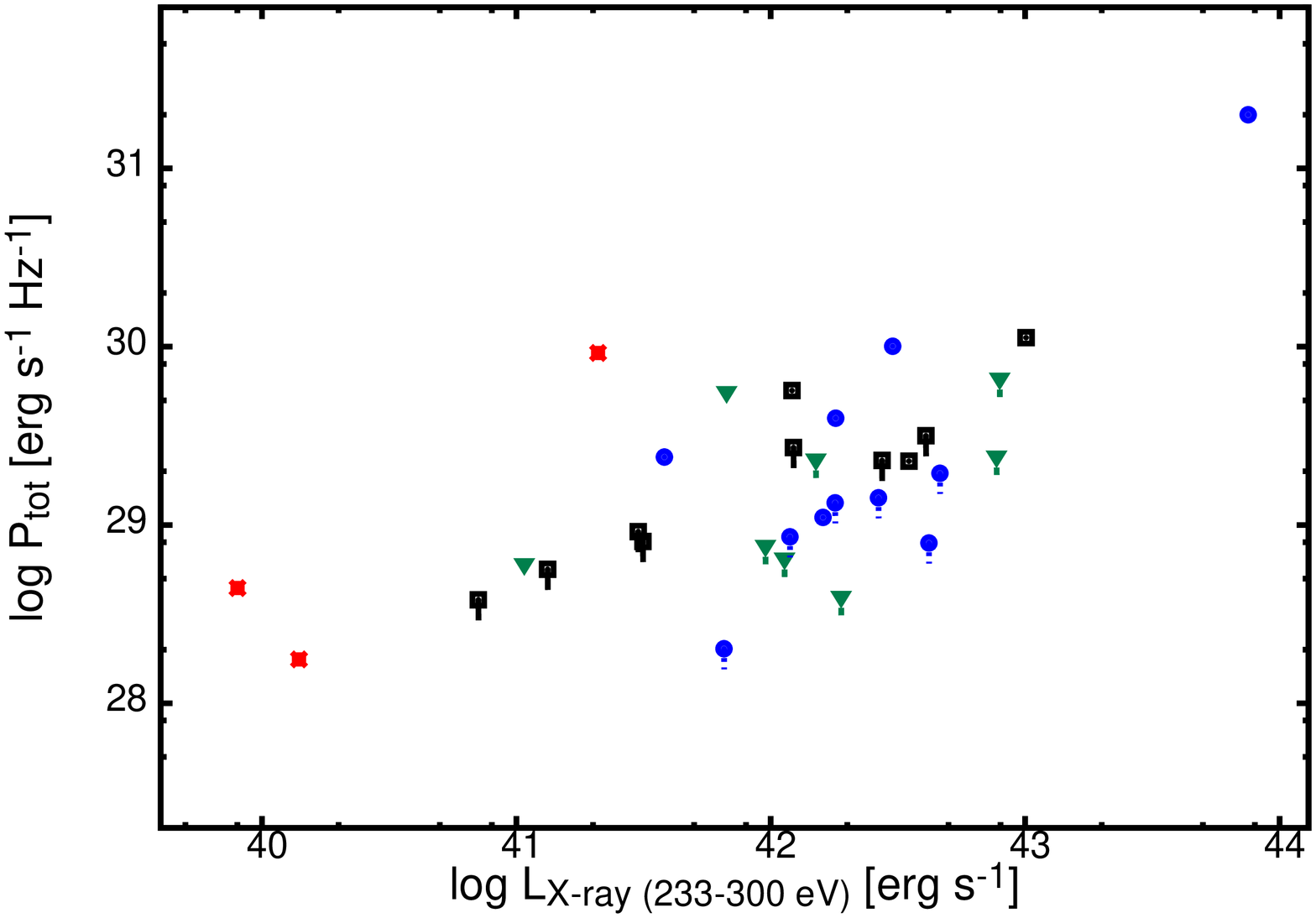} \hspace{2mm}
\includegraphics[angle=270,width=8.5cm]{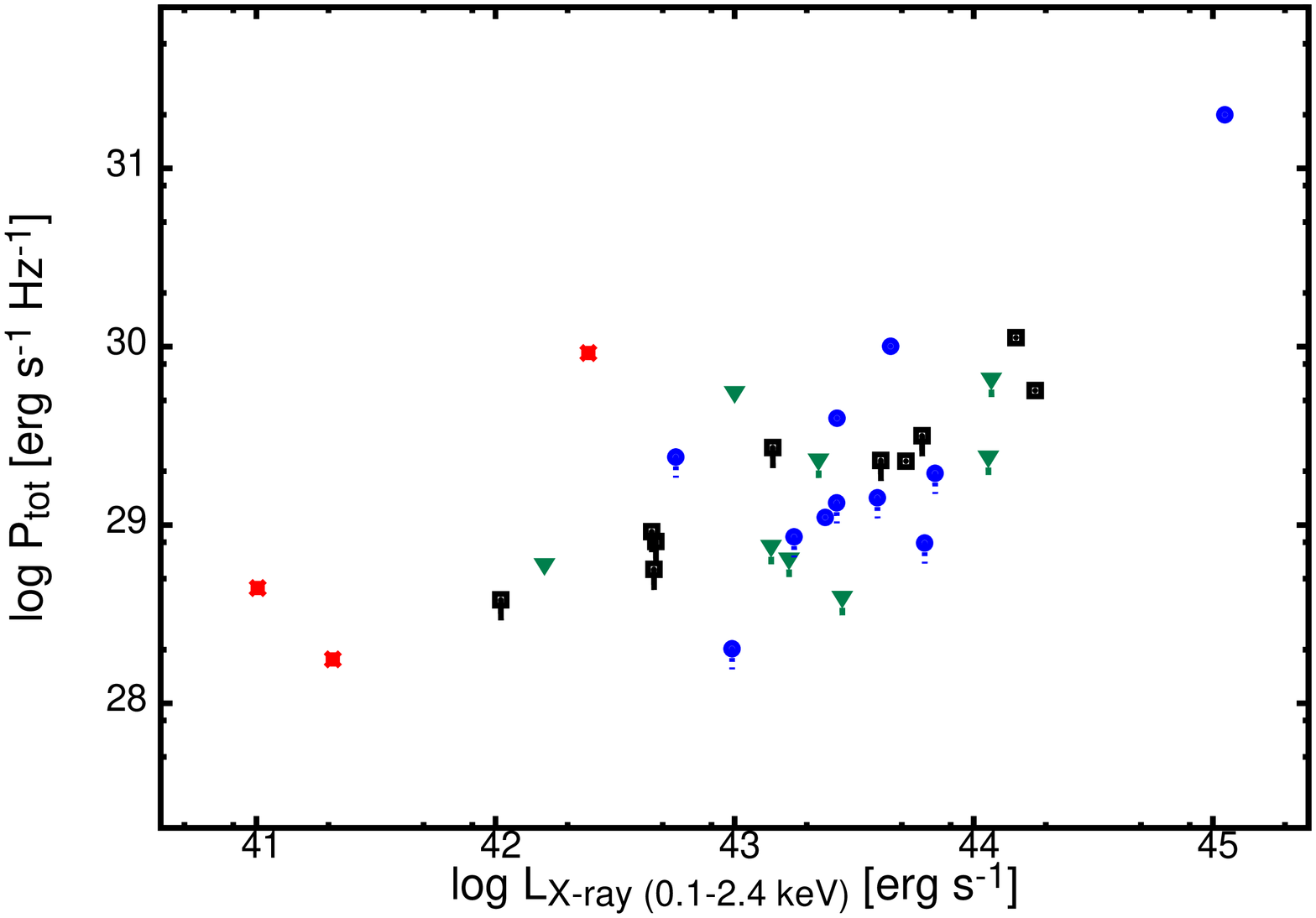}
\caption{Radio power at 1.4~GHz from VLA~$B$-array configuration
versus luminosities of
the soft X-ray band (233--300 eV) and in the hard X-ray band (0.1--2.4 keV).
We use softer spectrum ($\Gamma$ = 3.0) in the soft X-ray band and
harder spectrum ($\Gamma$ = 1.5) in the hard X-ray band (GMW09)
for NLS1 and Seyfert~1 galaxies with the assumption that they
have a softer spectrum in the soft X-ray band and harder spectrum
in the hard X-ray band.
Whereas Seyfert~2 galaxies have harder spectrum in both soft as well as
hard X-ray bands.
Point colours and styles are as in Figure~\ref{fe-o-h-corr},
and error bars are smaller than the size of the downward-pointing lines.}
\label{xray-corr}
\end{figure*}

\subsection{Correlations between Radio and Line Luminosities}

In Figure~\ref{fe-o-h-corr}, we plot correlations between radio power and luminosities of
[Fe\,X] (top-left panel),
[Fe\,XI] (top-right panel),
[Fe\,VII] (middle-left panel),
[O\,III] (middle-right panel),
[O\,I] (bottom-left panel), and
H$_{\alpha}$ (bottom-right panel.
[O\,III] and H$_{\alpha}$ line luminosities
correspond to combined core and wing components and combined narrow and
broad H$_{\alpha}$ model components, respectively (GMW09).
In Section 2.1, we discussed inherent biases in the sample
and hence possible difference between NLS1, Seyfert~1 and Seyfert~2 galaxies
are seen in H$_{\alpha}$ luminosity, X-ray luminosity, and ratio of [Fe\,X] and [Fe\,VII] lines.
Barring, correlation between radio power and
H$_{\alpha}$ luminosity, X-ray luminosity and ratio of [Fe\,X] and [Fe\,VII] lines,
which we discuss below (Section~\ref{rad-halp}, \ref{rad-xray},
\ref{rad-xray-rat}, respectively), none suggest that the distribution of
NLS1, Seyfert~1 and Seyfert~2 galaxies are dissimilar.
We therefore find no observational evidence against the unified
scheme hypothesis.  The K--S test gives a significance level of 0.031 or
better that the Seyfert types, Seyfert~1 galaxies along with NLS1 galaxies
and Seyfert~2 galaxies
are drawn from the same parent population,
and there is no significant differences between Seyfert~1 galaxies along with NLS1 galaxies
and Seyfert~2 galaxies, based on the distributions of
[Fe\,X], [Fe\,XI], [Fe\,VII], [O\,III], and [O\,I],
which was also independently concluded by GMW09.

Below we discuss those cases where we clearly see segregation between
the Seyfert types and their implications on the unified scheme
hypothesis.

\subsubsection{Radio and H$_{\alpha}$ Luminosities}
\label{rad-halp}

The radio luminosity versus H$_{\alpha}$ luminosity correlation plot in
Figure~\ref{fe-o-h-corr} (bottom-right panel) show that, both types
of Seyfert galaxies have similar radio powers, but Seyfert~2 galaxies have
systematically lower H$_{\alpha}$ luminosity than Seyfert~1 galaxies along with NLS1
galaxies.
the latter is because the broad component of H$_{\alpha}$
in Seyfert~2 galaxies is obscured by the obscuring torus
and is always not detected, which is more evident in Figure~A2b
(from supplementary material) of GMW09.
Since the sample is selected on FHIL emission,
lines with ionization potential $\gtrsim$100~eV (GMW09), their radio emission
is expected to have similar distributions of radio powers
within the framework of unified scheme hypothesis.
This is indeed the case, i.e., both types of Seyfert galaxies have similar radio powers,
consistent with the predictions of unified scheme.

\subsubsection{Radio and X-ray Luminosities}
\label{rad-xray}

There are only 32 sample sources with \textit{ROSAT} detections,
of which only three are Seyfert~2 galaxies,
which on average have lower X-ray luminosities but their radio powers
are similar to rest of the sample sources.
In Figure~\ref{xray-corr}, we show the correlation
between radio and soft X-ray luminosities fitted to Seyfert galaxy populations.
Unfortunately, \textit{Chandra}/\textit{XMM}-\textit{Newton} data
do not exist for the sample.
Seyfert~1 along with NLS1 galaxies and Seyfert~2 galaxies are known to
have, on average, steeper soft X-ray spectra and flatter hard X-ray spectra,
respectively
\citep{KCU1990,MasHessetal1994,Boller2000,Dadina2008}.
Here, we used single uniform photon index,
$\Gamma$ = 1.5 as well as steeper photon index ($\Gamma$ = 3.0)
data from GMW09 for Seyfert~1 along with NLS1 galaxies
to convert \textit{ROSAT} count rates to fluxes
\citep[GMW09;][]{Cappietal2006,Dadina2008}.
We find that the same correlation function applies between
radio and soft X-ray luminosities for all types of Seyfert galaxies
with no systematic differences between them.
We therefore conclude that there is no difference between radio luminosities of
NLS1, Seyfert~1 and Seyfert~2 galaxies, consistent with unified scheme.

\subsubsection{Radio Luminosity, and Ratio of\\[0pt] [Fe\,X] and [Fe\,VII] Lines}
\label{rad-xray-rat}

\begin{figure}[!t]
\centering
\includegraphics[angle=270,width=8.35cm]{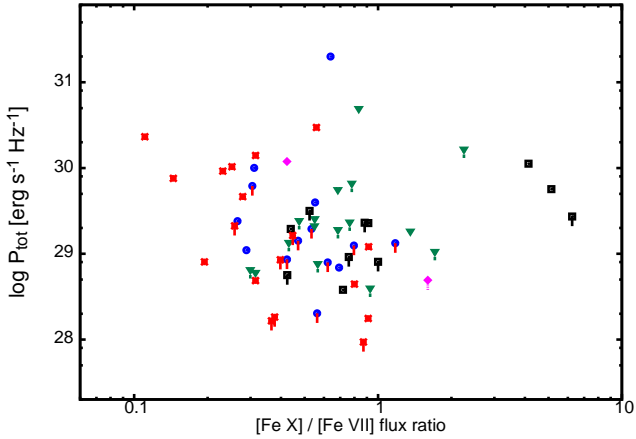}
\caption{Radio power at 1.4~GHz from VLA~$B$-array configuration
versus [Fe\,X]/[Fe\,VII] flux ratio.
Point colours and styles are as in Figure~\ref{fe-o-h-corr},
and error bars are smaller than the size of the downward-pointing lines.}
\label{flux-ratio-rad}
\end{figure}

Differences between the two types of Seyfert galaxies for
the ratio of [Fe\,X] and [Fe\,VII] provide evidence that in Seyfert~2 galaxies
there exist partially obscured [Fe\,X] emission and hence low ratio
of these two lines (GMW09).
This is consistent within the stratified wind model framework discussed below
(Section~\ref{core-det}), that
the [Fe\,X] is emitted on size scales that is comparable to the size of
the dusty molecular torus, and hence obscuration may be important.
As shown in Figure~\ref{flux-ratio-rad},
NLS1 galaxies, Seyfert~1 galaxies and Seyfert~2 galaxies show
difference in the ratio of [Fe\,X] and [Fe\,VII]
which are possibly because of (i) absorption of the [Fe\,X]
or (ii) different ionizing continuum (GMW09),
where the latter is perhaps due to the bias in the sample.
Briefly, in Seyfert~1 galaxies,
the line width of the [Fe\,X] line profiles is broader, and
the ratios of [Fe\,X] and [Fe\,VII] tend to be higher,
whereas it is in opposite sense for the Seyfert~2 galaxies (GMW09).
These facts suggest that [Fe\,X]-emitting clouds lie close to the radio
core and in certain lines of sights, some fraction of the [Fe\,X] flux
(i.e., the broad component) may be hidden by the obscuring torus (GMW09).
Hence, there is possibly an excess of ionizing photons from active nucleus and
we expect an over-production of [Fe\,X] or the higher ratio of
[Fe\,X] and [Fe\,VII] in the NLS1 galaxies and also in the Seyfert~1 galaxies,
whereas this would again be in opposite sense for Seyfert~2 galaxies
due to the obscured view of active nucleus by the obscuring torus.
These are in line with the expectations of the unified scheme hypothesis.
Hence, in Figure~\ref{flux-ratio-rad},
we find all types of Seyfert galaxies to have similar radio powers,
but the Seyfert~2 galaxies on average have less [Fe\,X] power per unit of
[Fe\,VII] power than the Seyfert~1 galaxies or NLS1 galaxies,
which is again consistent with the predictions of unified scheme.

\subsection{Starburst vs. AGN Nature of FHIL-Emitting Seyfert Nuclei}
\label{core-det}

We first investigate if the radio emission from 8\% radio-intermediate
sources in the sample been enhanced due to intense star formation.
Quantitatively, the mean radio power of radio-intermediate sources,
$P^{\rm tot}_{\rm 1.4\,GHz} \simeq 8.5 \times 10^{30}$
\lum\ \perhz\, implying a star formation rate (SFR)
of $\sim 220$ $M_{\odot}\, {\rm yr^{-1}}$ for stars massive enough to form
supernovae, i.e., $M \ge 5 M_{\odot}$.
\citet{Mainierietal2005,Vignalietal.2009} and \citet{Rosarioetal2012}
showed that such high rates of star formation are seen in high-$z$ AGNs,
but they are more than the SFR inferred for
(i) typical nearby Seyfert galaxies \citep{Alonso-Herrero}
by an order of magnitude \citep{Laletal2011},
and also for (ii) the most luminous AGNs in the SDSS \citep{Liuetal2009}.
Therefore, it seems that the enhanced radio emission in radio-intermediate sources
as compared to radio-quiet sources in the sample
is unlikely to be due to intense star formation.

Secondly,
the radio spectra of most Seyfert galaxies are known to be steep
power laws near 1.4\,GHz, $<$$\alpha$$>$ $\sim$0.8 \citep{Laletal2011}
indicating that the radio flux densities at this frequency
are dominated by non-thermal emission.
However, the strong optical emission lines imply the presence of ionized gas
that could contribute significant thermal emission at frequencies
$\ge$1.4 GHz.
It is thus important to compare the expected thermal emission and the
total radio flux densities observed from Seyfert galaxies in the sample.
We now test and demonstrate that both, radio-quiet
and radio-intermediate Seyfert galaxies in the sample have low-levels of
expected star formation rates, lower than
the limits that is needed to account for the observed radio emission.
Following \citet{Ho2005,Ho2008} and also \citet{LalHo2010},
the [O\,II]-derived SFRs are estimated to be
$\sim$0.2--10~M$_{\odot}\, {\rm yr^{-1}}$,
with a mean ($\approx$ median) value of $\sim$0.6~M$_{\odot}\, {\rm yr^{-1}}$
for the Seyfert galaxies in the sample;
we use the ratios of ionisation potentials between [O\,I], [O\,II] and [O\,III] emission lines
and [O\,I] and [O\,III] fluxes,
and thereby determine [O\,II] fluxes.
Using Equation 6 of \citet{Bell2003},
the predicted radio emission from SFR is
$\sim1.1\times 10^{28}$ \lum\ \perhz\ at 1.4 GHz,
which is at least an order of magnitude less than the mean radio luminosity
of the sample sources (Figure~\ref{xu-gel-samp} and \ref{fe-o-h-corr}).
Alternatively, assuming all lines of sight through the broad line region
are optically thick in the radio, an upper limit to the thermal
emission at the mean distance of $\sim$400~Mpc for objects in the sample
gives a flux density of $\lesssim$ 0.004 mJy, quite undetectable for
typical values of densities of protons and electrons,
$n_{\rm p} = n_{\rm e} \simeq 10^9$ cm$^{-3}$ in broad line clouds,
of size $\simeq$ 0.1~pc having a temperature $\simeq$ 1.5 $\times 10^4$ K
\citep{Koski1978},
a filling factor $\simeq$ 10$^{-4}$ \citep{UWS81}, and the H$\beta$ luminosity
equal to the [O\,III]\,$\lambda$\,5007\,\AA\ luminosity of 10$^{42}$ erg~s$^{-1}$
\citep{UWS81}.
On similar lines we calculate radio power $\lesssim 10^{27}$ erg~s$^{-1}$~Hz$^{-1}$,
using canonical $n_{\rm e} \simeq$ 10$^3$ cm$^{-3}$,
temperature $\simeq$ 1.5 $\times 10^4$ K, filling factor $\simeq$ a few times
10$^{-2}$ for narrow line clouds of size $\simeq$ 0.1--1.0~kpc
at mean distance for objects in the sample.
Clearly, it is the narrow line region that has relatively larger
thermal emission contribution
to the observed radio emission as compared to the broad line region.
But the derived thermal contributions at 1.4\,GHz are small,
at a level of 10\% or less of the observed flux density for objects
in the FHIL-emitting Seyfert galaxy sample.
Also, \citet{Guetal2006}
used the observed emission-line ratios along with population synthesis models
and concluded that although there is a contribution of starbursts to the
nuclear emission in a radio-quiet Seyfert galaxies,
the contribution of the hidden active nucleus always dominates.
Their conclusion compares well for the FHIL-emitting sources studied here.

Finally, in the light of a stratified wind model,
first proposed by \citet{Osterbrock1991},
outflowing photoionized clouds produce [Fe\,XI], [Fe\,X],
[Fe\,VII] and possibly [O\,III] (GMW09),
with high-ionization lines are produced preferentially
at small distances from the core,
while low-ionization lines are preferentially produced at larger radii.
More specifically
[Fe\,XI]- and [Fe\,X]-emitting clouds lie closest to the radio core and
are on the scale of broad-line region or of obscuring torus,
[Fe\,VII]-emitting clouds are more extended and may approach
the scale of narrow-line region, and
finally [O\,III]- and [O\,II]-emitting clouds,
which do not show a velocity shift with respect to [S\,II] (GMW09),
lie farthest from the radio core and more likely are the decelerated part
of the stratified wind outflow model \citep{Colbertetal1996a,Colbertetal1996b}.
If this model is correct, we expect to find a strong correlation between
emission-line shifts and ionization potential of the line-emitting species;
implying density and ionization stratification \citep{Komossaetal2008}.
Such correlations were indeed reported by GMW09
for some of the ion species.
Thus, there is a ionization stratification associated with clouds,
and possibly clouds on scales of narrow-line region, which
have relatively the least amount of ionization stratification,
have large optical depths at 1.4\,GHz.
These clouds are responsible for free--free absorption of radio emission
from the core, leading to inferred low radio detection rate irrespective of
Seyfert types, which is consistent with orientation based unified scheme hypothesis.

\section{Conclusions and Future Directions} \label{summary}

This paper presents radio properties of a sample of
61 sources containing a diverse range of Seyfert galaxies.  This is
one of the largest, homogeneous samples of Seyfert galaxies with strong FHIL
emission selected from the SDSS to date (GMW09).  Here we used high-resolution FIRST data
\citep[5{\asec} resolution images,][]{becketetal}
along with low-resolution NVSS radio data
\citep[45{\asec} resolution images,][]{Condonetal1998}, both at 1.4 GHz.
Our main results are:
\begin{enumerate}

\item{Our detection rate of 49\% at 1.4 GHz is lower than many other
Seyfert galaxy samples, except far-infrared selected sample compiled by \citet{royetal}.}

\item{A high fraction (76\%) of compact cores are seen within confirmed
detected sources.  The radio emission within these compact cores is confined
to physical sizes $\lesssim$8~kpc.
The remaining 24\% objects contain extended, core-jet structures,
typical of nearby Seyfert galaxies.}

\item{The detection rate of compact radio structure in NLS1, Seyfert 1
and Seyfert~2 galaxies is consistent with the unified scheme hypothesis.}

\item{The distributions of radio power for Seyfert~1 along with NLS1 galaxies and
Seyfert~2 galaxies for the sample are not significantly different.
This is possibly consistent with the unified scheme hypothesis.
However, given the size of the sample with several upper limits,
it is difficult to uncover any subtle differences that might exist between
the types of Seyfert galaxies.}

\item{There is possibly no evidence of relativistic beaming in nuclei for objects
in our sample.}

\item{Approximately 8\% of the sample sources have ratio of radio luminosity
and [O\,III]\,$\lambda$\,5007\,\AA\ luminosity such
that they qualify as radio-intermediate sources and the remaining are
radio-quiet.}

\item{The distributions of line luminosities and
the X-ray luminosities for the two Seyfert types are also consistent with
the unified scheme hypothesis.}

\item{These sample objects clearly show AGN activities with $\le$10\% contribution
from thermal emission, and they show poor \mbox{($\sim0.6$ \solmass\ \peryr)} SFRs,
typical of Seyfert galaxies.}

\item{It seems that there is ionization stratification associated with clouds,
and possibly clouds on scales of 0.1--1.0~kpc have large optical depths at 1.4\,GHz,
which are responsible for free--free absorption of radio emission from the core,
possibly, leading to low radio detection rate for these objects.}

\end{enumerate}

However, a weakness remains with the data; these survey images are
sensitive to extended emission from starburst activity around the nucleus
in addition to the compact emission with the Seyfert core.
Future deeper, high-resolution and high-sensitivity radio observations at both,
high and low radio frequencies are needed to
(i) test and understand the low detection rate,
(ii) resolve the FIRST survey images into structures of sizes less than a kiloparsec,
(iii) test the predictions of free--free absorption of radio core emission
by the narrow-line region clouds, and
(iv) understand if these objects have AGN-like, high-brightness
temperature flat-spectrum cores,
or these have steep-spectrum diffuse emission.


\acknowledgments

The author thanks the anonymous referee for suggestions and criticisms which improved
the paper.
He also thanks Prof. M. Elvis and Prof. M.J. Ward for many helpful conversations, and
Dr. M.J. Hardcastle and Dr. D.A. Green for a careful reading of this manuscript.
The VLA is operated by the US National Radio Astronomy Observatory which is
operated by Associated Universities, Inc., under cooperative agreement with
the National Science Foundation.  The National Radio Astronomy Observatory is
a facility of the National Science Foundation operated under cooperative
agreement by Associated Universities, Inc.  This research has made use of the
NASA/IPAC Extragalactic Database (NED) which is operated by the Jet Propulsion
Laboratory, California Institute of Technology, under contract with NASA.


\end{document}